\documentclass[preprintnumbers,secnumarabic,amsmath,amssymb,nofootinbib,floatfix]{revtex4}
\usepackage{varioref,exscale,latexsym,amsmath,amssymb}
\usepackage{graphicx}
\pdfoutput=1

\usepackage{graphicx}
\usepackage{slashed}
\usepackage{dcolumn}
\usepackage{bm}
\usepackage{hyperref}
\usepackage{color}
\usepackage{xcolor}
\newcommand{\beq}{\begin{equation}}
\newcommand{\eeq}{\end{equation}}
\newcommand{\bea}{\begin{eqnarray}}
\newcommand{\eea}{\end{eqnarray}}
\newcommand{\nn}{\nonumber}

\def\lsi{\raise0.3ex\hbox{$<$\kern-0.75em\raise-1.1ex\hbox{$\sim$}}}
\def\gsi{\raise0.3ex\hbox{$>$\kern-0.75em\raise-1.1ex\hbox{$\sim$}}}

\def\beq{\begin{equation}}

\def\eeq{\end{equation}}

\def\beqa{\begin{eqnarray}}

\def\eeqa{\end{eqnarray}}

\begin{document}
\preprint{ACFI-T19-08}

\title{{\bf  Unitarity, stability and loops of unstable ghosts}}

\medskip\

\medskip\

\author{John F. Donoghue}
\email{donoghue@physics.umass.edu}
\affiliation{~\\
Department of Physics,
University of Massachusetts\\
Amherst, MA  01003, USA\\
 }

\author{Gabriel Menezes}
\email{gabrielmenezes@ufrrj.br}
\affiliation{~\\
Department of Physics,
University of Massachusetts\\
Amherst, MA  01003, USA\\
}

\affiliation{~ Departamento de F\'{i}sica, Universidade Federal Rural do Rio de Janeiro, 23897-000, Serop\'{e}dica, RJ, Brazil \\
 }

\begin{abstract}
We present a new understanding of the unstable ghost-like resonance which appears in theories such as quadratic gravity and Lee-Wick type theories. Quantum corrections make this resonance unstable, such that it does not appear in the asymptotic spectrum. We prove that these theories are unitary to all orders. Unitarity is satisfied by the inclusion of only cuts from stable states in the unitarity sum. This removes the need to consider this as a ghost state in the unitarity sum.  However, we often use a narrow-width approximation where we do include cuts through unstable states, and ignore cuts through the stable decay products. If we do this with the unstable ghost resonance at one loop, we get the correct answer only by using a contour which was originally defined by Lee and Wick. The quantum effects also provide damping in both the Feynman and the retarded propagators, leading to stability under perturbations.
\end{abstract}

\maketitle

\section{Introduction}

Theories such as quadratic gravity \cite{Stelle:1976gc, Julve:1978xn, Fradkin:1981hx, Tomboulis, Salvio:2018crh, Donoghue:2018izj, Einhorn, Strumia, Donoghue, DM, Holdom, Mannheim, Hooft, Shapiro, Smilga, Narain, Anselmi, bos, Alvarez-Gaume} and Lee-Wick theories \cite{Lee:1969fy, Lee:1969fz, Lee:1970iw, Cutkosky:1969fq, Coleman, Grinstein:2007mp, Grinstein:2008bg, Modesto:16, Accioly:17} have propagators which contain both
quadratic and quartic momentum dependence.
In addition to the pole at $q^2 =0$, this combination will produce a high mass pole, via
\begin{equation}\label{partialfrac}
  \frac{1}{q^2 - \frac{q^4}{\mu^2}} =  \frac{1}{q^2} -  \frac{1}{q^2 - {\mu^2}} \ \ .
\end{equation}
The negative sign for the second term implies that this pole is ghost-like, i.e. with negative norm. One therefore expects trouble with some combination of unitarity, stability, causality, etc. in such a theory. However, as Lee and Wick originally noted, interactions in such a theory make the heavy state unstable, with a width which can be calculated in perturbation theory. This feature is a crucial modification as it removes the ghost from the asymptotic spectrum. Past experience with Lee-Wick theories indicates that they can be stable and unitary, although causality does seem to be violated on microscopic scales of order the width of the resonance \cite{Lee:1969fy, Coleman, Grinstein:2008bg, DMcausality}. In this paper we describe further the unitarity and stability of such theories and come to an understanding of how unitarity is satisfied in the presence of unstable ghosts.

\subsection{Unitarity with normal resonances}

Unitarity describes the conservation of probability for the $S$-matrix. It states
\beq
\langle f |T| i \rangle - \langle f |T^\dagger| i \rangle = i \sum_j \langle f |T^\dagger| j \rangle
\langle j |T| i \rangle
\label{unit}
\eeq
where we used the definition of the transfer matrix $T$, namely $S = {\bf 1} + i T$. Here the associated states are the asymptotic single and multiparticle states of the theory. In processes that involve loop diagrams, the sum over real intermediate states can by accomplished by the Cutkosky cutting rules \cite{Cutkosky:1960sp} which project out the on-shell states.

Procedurally we often look first at the free field theory to identify the free particles. Then when we include interactions, some of these particles become unstable and no longer appear as the asymptotic states of the theory. As far as the S-matrix is concerned, this is a significant change. The particles were originally needed in the Hilbert space for completeness, but then are no longer present in the interacting theory. The question then arises of how to treat such unstable particles in unitarity relations. Should one include them in the sums over states required for unitarity?

The answer was provided by Veltman in $1963$ \cite{Veltman:1963}, see also \cite{Veltman:74, Rodenburg, Lang, Denner:2014zga}. He showed that unitarity is indeed satisfied by the inclusion of {\em only} the asymptotically stable states. Cuts are not to be taken through the unstable particles, and unstable particles are not to be included in unitarity sums.

However, there is a corollary which is useful in practice. In the narrow-width approximation, where the coupling to the decay products is taken to be very small, the off-resonance production becomes small and only resonance production is important. As we demonstrate in Sec. \ref{NWA}, in this limit a cut taken through the unstable particle with its width set to zero reproduces the same result as a cut through the decay products.

This combination reinforces our intuition. The full calculation only requires the stable states, as unitarity demands. But when particles are nearly stable, we may approximate them as being stable in practical calculations.

\subsection{Unstable ghost-like resonances}

The ghost-like resonance in Lee-Wick type theories should also not be treated as an asymptotic state. In Sec.~\ref{Veltman} we will demonstrate how the Veltman's procedures can be employed for the calculation of cuts in diagrams involving such unstable ghost-like states, and show that the usual cuts are obtained on the stable particles. Unitarity is satisfied for such cuts, and no cuts should be taken for the unstable ghost. There is no need for a modified contour in calculating the discontinuity, and the momentum integration runs along the real axis just above the cut.

However a slight difference arises in defining the narrow-width approximation. While all the stable cuts are treated with the usual $i\epsilon$ prescription and are located below the integration contour, the unstable ghost carries a finite width and appears above the standard contour. If we wish to treat the unstable ghost as if it were a stable particle in the narrow-width approximation, we show that the correct answer is reproduced if and only if the contour is defined to pass above the location of the unstable pole. This is the origin of the Lee-Wick contour which was a somewhat puzzling feature of such theories. It is only invoked to reproduce the correct result of unitarity cuts through the stable particles.

\subsection{Why this matters}

In path-integral treatments, the stable states of a theory can generally be identified through the poles in the propagators. These then form the Hilbert space of the theory and are relevant for the unitarity relations. In the theories which we are discussing, it is relatively easy to identify that the ghost-like degrees of freedom are not part of the asymptotic spectrum.

In contrast, using canonical quantization we traditionally consider the free field theory first. When trying to treat ghosts as regular free particles, we need to develop new quantization procedures. This task has occupied much of the literature on Lee-Wick type theories. Indeed such procedures can be defined, although they appear somewhat unusual,  and involve new concepts like indefinite metric or PT quantization \cite{Lee:1969fy, Bender:2007wu, Salvio:2015gsi, Raidal:2016wop}. One then debates whether to include such states in the unitarity sum or to define the path integration over such ghost variables.

While it is perhaps reassuring that canonical quantization is possible, it is an unnecessary step. Since these effects do not appear in the physical spectrum, we do not need to canonically quantize the ghosts. The path integral is over the physical field only and the resonance appears as a calculable dynamical effect in the propagator. The resonance will have physical effects, but we do not need to describe its free field quantization. Only stable states appear in the unitarity sum and their cutting rules will be the standard ones.
Our proof of unitarity in the presence of unstable resonances allows us to consider these theories further without discussing the canonical quantization of the unstable ghosts.

\section{Ghost resonances}

The theories which we are studying have propagators of the form
\beq\label{propagator}
iD(q) = \frac{i}{q^2 +i \epsilon - \frac{q^4}{M^2} + \Sigma (q) }  \ \  .
\eeq
The pole at $q^2=0$ is the stable particle of the theory. The function $\Sigma (q)$ is the self-energy or vacuum polarization function. At high energies, it picks up an imaginary part due to coupling to light particles. At one-loop order, this typically has the form
\beq
\Sigma (q)= - \frac{\gamma}{\pi} \log \left(\frac{-q^2-i\epsilon}{\mu^2}\right) =  \left[-\frac{\gamma}{\pi} \log \left(\frac{|q^2|}{\mu^2}\right) +i\gamma \theta(q^2) \right]
\eeq
for some calculable quantity $\gamma$ with dimensions of mass squared. In general $\gamma$ can be a function of the momentum, and in quadratic gravity $\gamma \sim Gq^4$. In our presentation here, for clarity we will neglect any masses of stable particles and the particles which appear in the vacuum polarization loop. It would be simple to restore any such masses.

The quartic momentum dependence is the novel feature in this propagator. It arises in theories which have four derivatives in the fundamental Lagrangian. We have chosen a particular sign for the quartic momentum dependence. The opposite sign would produce a tachyonic pole on the space-like real axis, which appears to be fatal for the theory. With the sign shown in Eq. \ref{propagator}, there is a massive resonance for timelike values of $q^2$. Expanding near that resonance, we get the form
\beq
iD(q) \bigg|_{q^2\sim m^2} \sim \frac{-i}{q^2-m^2 -i \gamma}  \ \   .
\eeq
The minus sign in the numerator tells us that this is a ghost-like resonance - it carries the opposite sign from the usual case. Equally important is that the width also comes with the opposite sign from normal. The resonance parameters are then $m_R^2 = m^2+i\gamma$.

The point here is that the normal resonances and ghost-like resonances have the structure
\beq\label{twosigns}
iD(q) \sim \frac{Zi}{q^2-m^2 +iZ\gamma}
\eeq
with $Z=+1$ for a normal resonance and $Z=-1$ for the ghost resonance. The change of the two signs together is important. In particular, the imaginary part of the propagator is independent of $Z$,
\beq
\textrm{Im}[D(q)] \sim \frac{-\gamma}{(q^2-m^2)^2 + \gamma^2}  \ \ .
\eeq
We will show in subsecton \ref{generic} that all viable (i.e. non-tachyonic) theories with quartic momentum dependence falls into this class.

Note also that the stable particle and the resonance appear in the same propagator, Eq. \ref{propagator}, so that we need not be talking about two separate particles, but rather two features in the same propagator. However, because of the partial fraction relation, Eq. \ref{partialfrac}, it is equally possible to separate the propagator into the equivalent of two particles. It is also generally possible to do this by a field redefinition at the level of the Lagrangian.

\subsection{Quadratic gravity}

In quadratic gravity, the interesting case involves the spin-two propagator. The general Lagrangian has the form

\begin{equation}\label{quadraticorder}
S_{\textrm{quad}} = \int d^4x \sqrt{-g}
\left[ \frac{2}{\kappa^2} R+\frac{1}{6 f_0^2} R^2  -\frac{1}{\xi^2}\left(R_{\mu\nu}R^{\mu\nu}
- \frac13 R^2\right)\right] \ \ .
\end{equation}
Here $\kappa^2 = 32\pi G$. We have dropped the cosmological constant and will be expanding about Minkowski spacetime, $g_{\mu\nu}= \eta_{\mu\nu}+ \kappa h_{\mu\nu}$. For more detail on this theory, see for instance~\cite{Stelle:1976gc,Salvio:2018crh,Donoghue:2018izj}.

The spin-two propagator only depends on $\kappa$ and $\xi$.
The vacuum polarization diagram adds a logarithmic term to the inverse propagator that is given by
\beq
D_2^{-1}(q) = \cdots
- \frac{\kappa^2 q^4 N_{\textrm{eff}}}{640\pi^2} \ln \left(\frac{-q^2-i\epsilon}{\mu^2}\right)
\eeq
where $ N_{\textrm{eff}}$ is contribution of the light matter fields and normal gravitons, normalized to the contribution of massless vector fields. For timelike momenta, it generates an imaginary part
\beq
\textrm{Im}[D_2^{-1}(q)] =  + \frac{\kappa^2q^4 N_{\textrm{eff}}}{640\pi} \theta(q^2)
\eeq
This generates the width for the ghost state. Our previous paper \cite{Donoghue:2018izj} on quadratic gravity has a detailed discussion of this feature.

\begin{figure}[htb]
\begin{center}
\includegraphics[height=60mm,width=65mm]{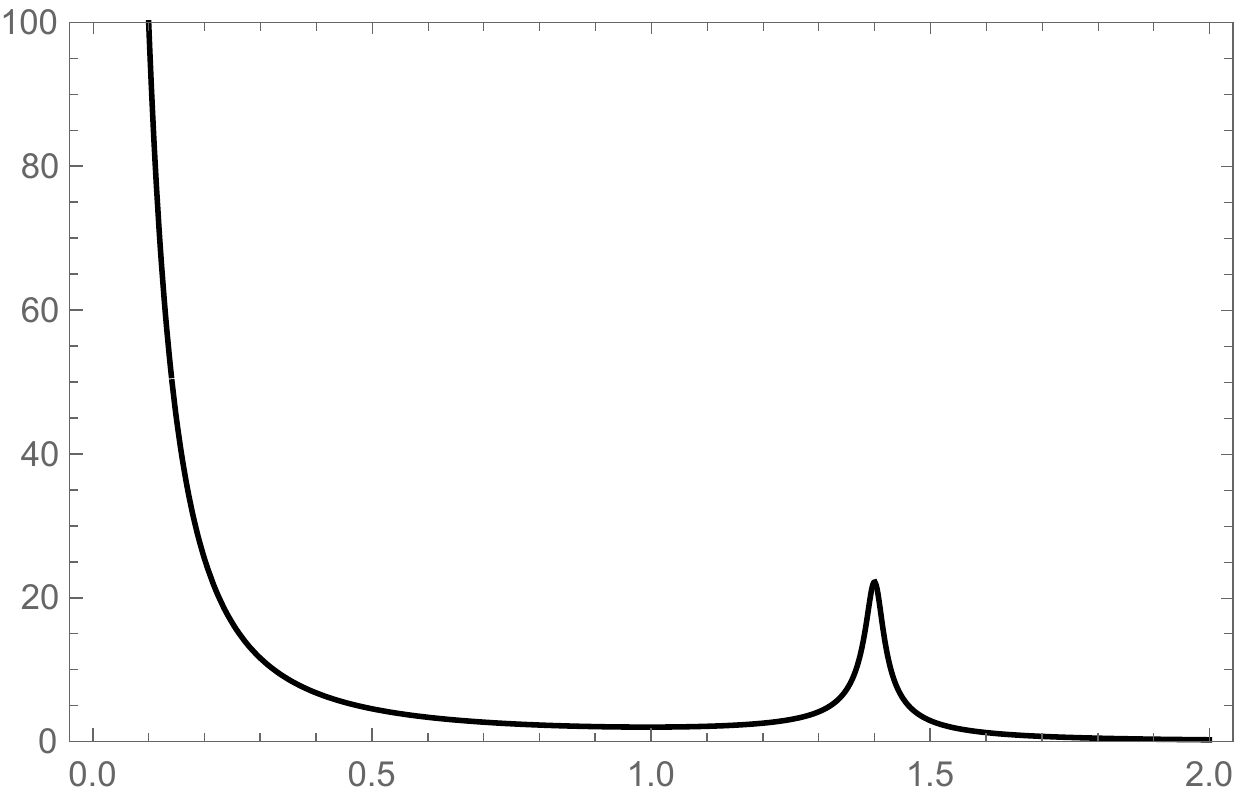}
\caption{The absolute value of the spin-two propagator for $\xi^2=1$, showing the high mass resonance. The x-axis is the momentum $|q|$ in the time-like region, in units where $\kappa=1$. The imaginary parts have been calculated with loops of Standard Model particles and gravitons. }
\label{grec}
\end{center}
\end{figure}

The overall scalar part of the spin-two propagator is
\beq\label{spintwo}
D_2(q) =  \left\{ {q^2+i\epsilon}- \frac{\kappa^2 q^4}{2\xi^2(\mu)}
- \frac{\kappa^2 q^4 N_{\textrm{eff}}}{640\pi^2} \left[\ln \left(\frac{|q^2|}{\mu^2}\right)-i\pi \theta(q^2)\right]
\right\}^{-1}
\eeq
for real values of the four-momentum. This is shown for timelike momenta in Fig. \ref{grec}. The unstable ghost resonance is visible. The peak occurs at momentum
\beq
q_r^2 = m_r^2 = \frac{2\xi^2(\mu)}{\kappa^2} + \frac{\xi^2(\mu)N_{\textrm{eff}}}{320\pi^2\kappa^2} \ln \left(\frac{|m_r^2|}{\mu^2}\right) = \frac{2\xi^2(m_r)}{\kappa^2}
\eeq
Evaluating the imaginary part at this momentum, one finds that the pole close to the real axis is located at
\beq\label{pole}
M^2 = m_r^2 + i \gamma~~~~~~~~~~~~~~~~~~~\gamma = {\xi^2m_r^2} \frac{N_{\textrm{eff}}}{320\pi}
\eeq
valid for weak coupling. Weak coupling $\xi<<1$ implies that $m_r$ is smaller than the Planck mass, and that the width is narrow compared to the mass. The residue at the pole is $-1$ in weak coupling. For large values of $\xi$ the solutions for the mass, width and residue must be found numerically.

\subsection{Lee-Wick QED}

Lee and Wick initiated a finite theory of QED, roughly by considering the Pauli-Villars regulator as a dynamical field. Because it enters with a minus sign, it also is ghost-like. The photon propagator in this case becomes
\beq
i D_{F\mu\nu}(q^2)= -i g_{\mu\nu} D(q^2)
\eeq
with
\beq \label{basicform}
D(q^2) =\frac{1}{(q^2+i\epsilon)\left[ 1 + \hat{\Pi}(q^2) - \frac{q^2}{\Lambda^2}\right]}
\eeq
with
\beq
\hat{\Pi}(q^2) = q^2 \frac{\alpha}{3\pi}\int_{4m_f^2}^{\infty} ds \frac{1}{s(s-q^2-i\epsilon)} \sqrt{1-\frac{4m_f^2}{s}}\left(1+\frac{2m_f^2}{s}\right)
\eeq
Here $\hat{\Pi}(q^2)$ is the finite part of the vacuum polarization function for a fermion of mass $m_f$, written in a dispersion theory representation. At large $q^2$, the vacuum polarization becomes logarithmic. Near the resonance, it is more convenient to renormalize the theory using a running coupling. In this case, the result is
\beq
\alpha D(q^2) \sim \frac{\alpha (\Lambda)}{q^2 \left[1 -\frac{\alpha}{3\pi} \log (q^2/\Lambda^2)  - \frac{q^2}{\Lambda^2}  + i \frac{\alpha}{3}\right] }
\eeq
to first order in the coupling. The large logarithm has been absorbed into the coupling and the remaining logarithmic running is weak near the pole.

This has the structure of our basic propagator, with $\gamma \sim \alpha \Lambda^2/3$.

\subsection{Generic quartic propagators}\label{generic}

More generally, our work applies to other theories with four derivatives in the Lagrangian. Consider for example a scalar field $\phi$ which is coupled to other light fields $\chi$, with
\beq\label{scalar}
{\cal L} = \frac12 \left[\partial_\mu \phi \partial^\mu \phi -m^2\phi^2\right] - \frac{1}{2M^2} \Box \phi \Box  \phi  - \frac{g}{2}\phi \sum_i \chi_i^2
\eeq
for $N$ light $\chi_i$ fields. The self-energy in this case proceeds through loops of the $\chi$ fields. It produces a residual self energy after mass renormalization
\beq
\Sigma = -\frac{Ng^2}{32\pi^2}\left[ \log \left(|q^2|/M^2\right)-i\pi\theta(q^2)\right]
\eeq
This matches our general form with $\gamma = Ng^2/(32\pi)$ if we neglect the mass of the $\phi$ field.

A slightly better variant, which can serve as a good scalar proxy for quadratic gravity, is defined by the Lagrangian
\beq\label{scalargravity}
{\cal L} = \frac12 \partial_\mu \phi \partial^\mu \phi  - \frac{\kappa^2}{4\xi^2} \Box \phi \Box  \phi  - \frac{\kappa}{2}(\Box \phi)\left( \phi^2 + \sum_i  \chi_i^2 \right)
\eeq
This variant does not need a mass for renomalization, because the interaction terms vanish on-shell using the massless equations of motion. In addition the momentum-dependent interaction mimics the effect of the two derivatives of the gravitational interaction. Here the residual self energy is
\beq
\Sigma = -\frac{\kappa^2 q^4(N+1)}{32\pi^2}\left[ \log \left(|q^2|/M^2\right)-i\pi\theta{q^2}\right]
\eeq
Indeed the parameters have been chosen such that the resulting propagator is identically equal to the spin-two propagator of quadratic gravity, Eq. \ref{spintwo}, with the identification of $N_{\textrm{eff}} = 20(N+1)$. We will use this variation below in order to avoid the tensorial complications of real gravity.

We can take this last model and add an auxiliary field in order to accomplish at the Lagrangian level the factorization of the propagator that one sees using partial fraction relations. To do this we introduce the auxiliary field $\eta$, using the Lagrangian
\beq\label{scalargravity2}
{\cal L} = \frac12 \partial_\mu \phi \partial^\mu \phi  -  \Box \phi \eta +\frac{\xi^2}{\kappa^2} \eta^2 - \frac{\kappa}{2}(\Box \phi)\left( \phi^2 + \sum_i  \chi_i^2 \right)
\eeq
Integrating out $\eta$ returns us to our original Lagrangian, Eq. \ref{scalargravity}. Now if we perform a field redefinition $\phi = h-\eta$, a little algebra turns this into
\begin{eqnarray}\label{scalargravity3}
{\cal L} &=& \left[\frac12 \partial_\mu h \partial^\mu h  - \frac{\kappa}{2}\Box h  \sum_i  \chi_i^2 \right] \nonumber \\
&-&  \left[\frac12 \left(\partial_\mu \eta \partial^\mu \eta -\frac{2\xi^2}{\kappa^2}\eta^2\right) - \frac{\kappa}{2}\Box \eta  \sum_i  \chi_i^2 \right] \nonumber \\
&-&  \frac{\kappa}{2}(\Box h -\Box\eta)( h-\eta)^2
\end{eqnarray}
Note in particular the overall minus sign in the second line.

From these examples, one is able to see that {\em all} theories with four derivative kinetic energies, $~\sim \Box \phi \Box \phi $ will fall into the class of theories which we are discussing, as long as one avoids the tachyonic pole at space-like momenta. The logic is as follows. Ordinary resonances arise when there is a coupling to the light states of the theory. The structure of the resonance propagator
is
\beq
iD_r(q) = \frac{i}{q^2 - m^2 + \Sigma(q)}   \  \ .
\eeq
The imaginary part of the self energy must be positive, that is
\beq
\textrm{Im}[\Sigma(q)] = \gamma(q),~~\theta(q^2) >0
\eeq
such that the resonance mass $m^2-\textrm{Re}[\Sigma] - i \textrm{Im}[\Sigma] = (M-i\Gamma/2)^2$. Now if the Lagrangian is modified with a $\Box^2 $ term, the propagator gets modified to be
\beq\label{generalcase}
iD(q) = \frac{i}{q^2 - m^2 + \Sigma(q)-q^4/\Lambda^2}
\eeq
where the sign on the new term has been chosen to avoid the tachyonic pole. If we set $\Lambda \to \infty$, we get a normal resonance where the near the pole the propagator has the structure given in Eq. \ref{twosigns} with $Z=+1$. However for large finite $\Lambda$
there is inevitably  a high mass resonance, when
$q^2 \sim \Lambda^2 $. For illustration we can neglect $m^2$ and $\textrm{Re}[\Sigma]$, and look at the structure near this resonance, such that
\beq
iD(q) = \frac{i}{q^2 - \frac{q^4}{\Lambda^2} + i \gamma(q) } = \frac{i}{\frac{q^2}{\Lambda^2}[\Lambda^2 - q^2 + i \gamma(q) (\Lambda^2/q^2)]} \sim \frac{-i}{q^2-\Lambda^2 -i \gamma}
\eeq
The residue at this pole is always negative - it is ghost-like. In addition, the sign of the width is always opposite from normal. That is, we find the correlated negative signs which we described in Eq. \ref{twosigns} with $Z=-1$. Indeed, for both finite $m$ and $\Lambda$, there will be resonances of both types contained in the same propagator. In both cases, the imaginary part of the self-energy arising from the coupling to stable states is the same, yet it manifests itself differently near the resonance because of the sign of the $q^4$ term.

\section{Stability and energy flow}\label{stability}

Because of the change in sign in front of the width in the denominator of the propagator, one might worry that there are exponentially growing modes. We show here that this is not the case. The stability of quadratic gravity has also been addressed using the equations of motion in curved backgrounds, although without including the decay width. The conclusion of  \cite{Chapiro:2019wua} has been that the theory is stable in a curved background, at least for curvatures that are below the ghost mass.  In \cite{Salvio:2019ewf} this was extended as at least metastability to curvatures beyond the scale of the ghost mass in the weakly coupled limit.  Our treatment is somewhat different, remaining in Minkowski space but including the very important effect coming from the fact that the ghost decays.

An important feature also emerges from this analysis. The energy flow associated with the ghost-like terms in the propagator is different from the usual case. What we normally refer to as ``positive energy'' is seen to be propagating backwards in time rather than the usual forward propagation \footnote{In \cite{DMcausality} we refer to these as ``Merlin modes'' in honor of the wizard in the Authurian tales  who ages backwards in time.}. Indeed the ghost propagator is the time reversed propagator of a normal resonance.

Let us study the coordinate space time dependence of the propagator by taking the Fourier transform, using the scalar part of the spin-two graviton propagator as our example
\beq
i D_2(x) = \int \frac{d^4q}{(2\pi)^4} ~e^{-iq\cdot x} i D_2(q)   \ \ .
\eeq
A little algebra allows us to write
\beq \label{exact}
D_2(q) = \left[\frac{1}{q^2} -\frac{1}{q^2 - \mu^2(q)}\right]
\eeq
without any approximation
\beq
\mu^2(q) = \frac{m_r^4\left[m^2_r+i\gamma \theta(q^2)+(\gamma/\pi)\log (|q^2|/m_r^2)\right]}{\left(m_r^2+ \frac{\gamma}{\pi}\ln (|q^2|/m_r^2)\right)^2 + \left({\gamma}\theta(q^2)\right)^2}   \ \ .
\eeq
In the weak coupling limit, $\gamma$ is a small number, so that we may approximate this by dropping the logarithmic term and keeping only the leading imaginary part. In this case the propagator becomes
\beq \label{approx}
D_2(q) = \left[\frac{1}{q^2} -\frac{1}{q^2 - m_r^2 - i\gamma \theta(q^2)}\right]  \ \ .
\eeq
\begin{figure}[htb]
\begin{center}
\includegraphics[height=80mm,width=80mm]{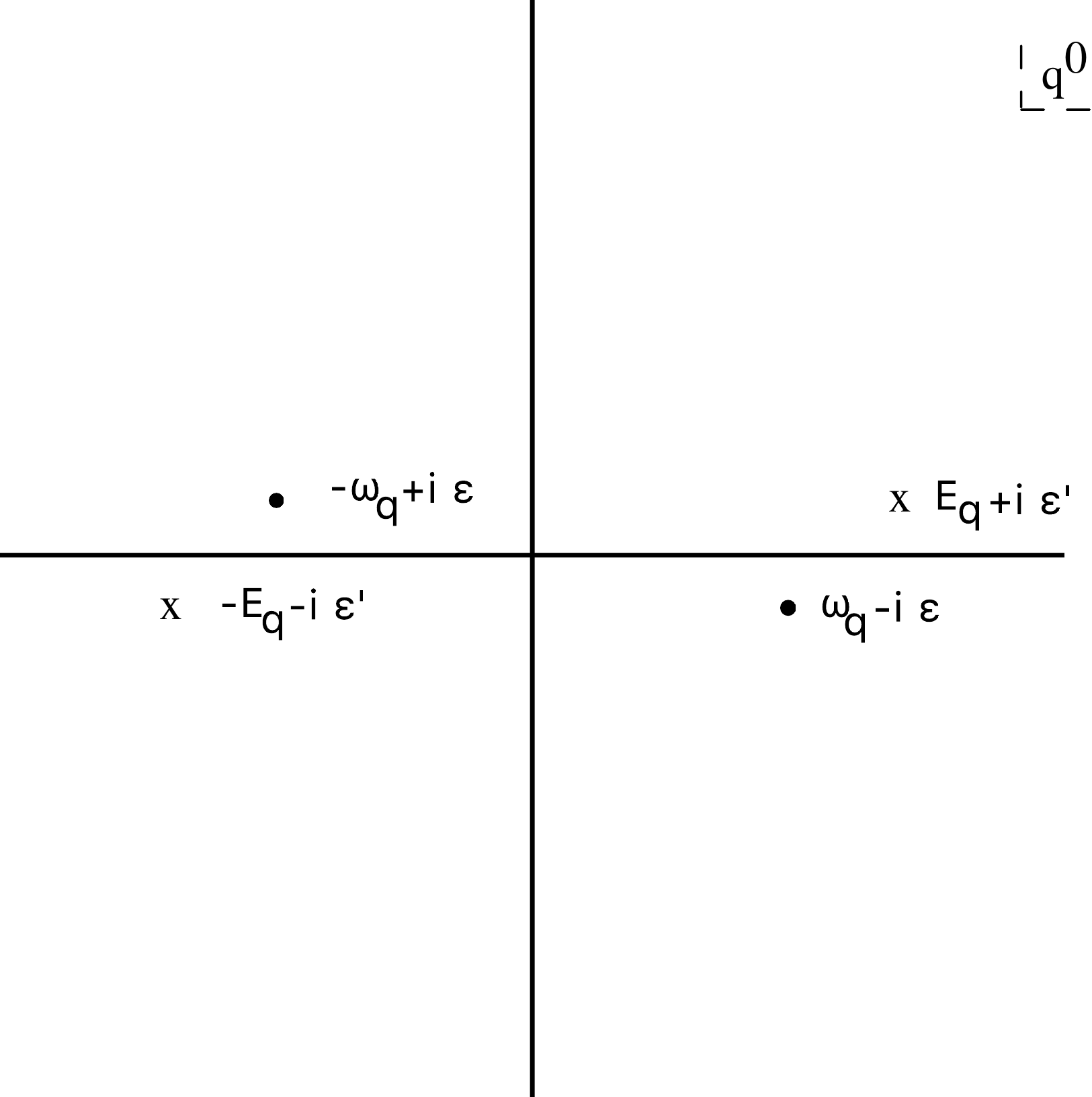}
\caption{The location of poles in the complex $q_0$ plane for the Feynman propagator. }
\label{contour}
\end{center}
\end{figure}

The propagator can be broken up into terms with forward propagation in time and backwards,
\beq
D_2(t, \vec{x}) = \Theta (t)D_{\textrm{for}} (x) + \Theta (-t) D_{\textrm{back}}(x)
\eeq
with $x_0=t$.
First consider $t>0$. In doing the $q_0$ integral, we need to close the contour in the lower half plane in order to have the contour at infinity give a vanishing contribution. There is the massless pole at
\beq
q_0^2 - \vec{q}^2 + i\epsilon =0
\eeq
which corresponds to $q_0 = \pm(\omega_q - i\epsilon)$ with $\omega_q=|\vec{q}|$.
There are massive poles at
\beq
q_0^2 - \vec{q}^2 - m_r^2 -i\gamma = 0
\eeq
or
\beq
q_0 = \pm \sqrt{E_q^2 +i\gamma} \sim \pm\left[E_q +i
\frac{\gamma}{2E_q}\right]
\eeq
with $E_q = \sqrt{\vec{q}^2+m_r^2}$, which produces the pole structure shown in Fig. \ref{contour}.
Performing the contour integral, we pick up the poles below the real axis.
This yields the forward propagator
\beq
D_{\textrm{for}}(t,\vec{x}) = -i\int \frac{d^3q}{(2\pi)^3 }\left[\frac{e^{-i(\omega_q t -\vec{q}\cdot\vec{x})}}{2\omega_q } - \frac{e^{i(E_q t -\vec{q}\cdot\vec{x})}}{2(E_q +i\frac{\gamma}{2E_q})} e^{-\frac{\gamma t}{2E_q}}\right]
\eeq
which shows the decaying exponential for the massive term, with the identification
\beq\label{width}
\gamma = m_r\Gamma \ \ .
\eeq
Also we note that the energy flow of the two terms is different, with the massless pole carrying what we normally call positive energy, while the massive resonance carries negative energy. The term describing propagation backwards in time is obtained for $t<0$ by closing in the upper half plane, with the result
\beq
D_{\textrm{back}}(t,\vec{x}) = -i\int \frac{d^3q}{(2\pi)^3 }\left[\frac{e^{i(\omega_q t -\vec{q}\cdot\vec{x})}}{2\omega_q } - \frac{e^{-i(E_q t -\vec{q}\cdot\vec{x})}}{2(E_q +i\frac{\gamma}{2E_q})} e^{-\frac{\gamma |t|}{2E_q}}\right]
\eeq
Again we see exponential decay, and the reversal of the energy flow between the two terms. We emphasize here that we are employing the usual contour for the propagator; we did not resort to any modified contour in order to obtain the above results.

\begin{figure}[htb]
\begin{center}
\includegraphics[height=80mm,width=80mm]{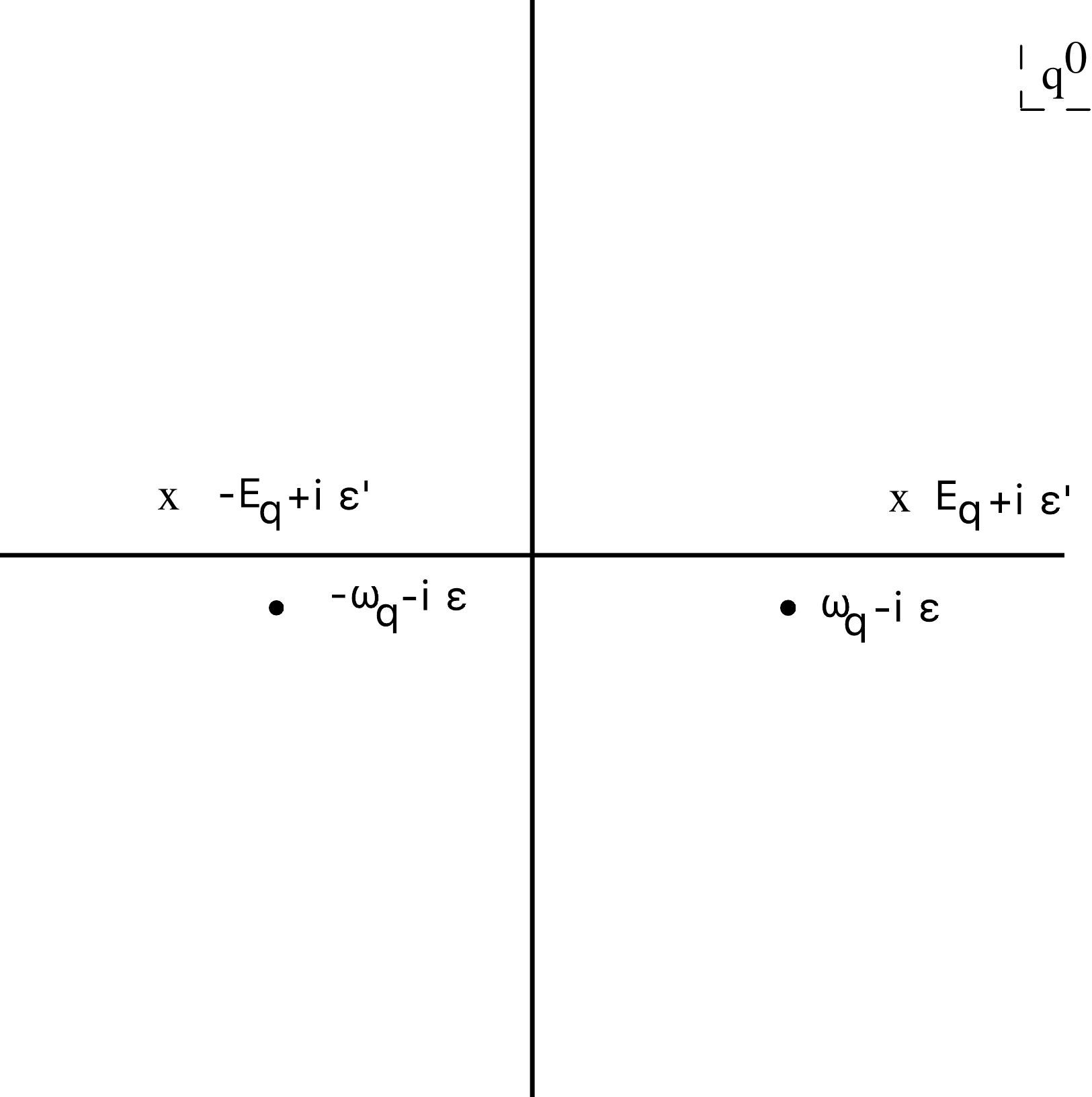}
\caption{The location of poles in the complex $q_0$ plane for the retarded Green function.}
\label{contour2}
\end{center}
\end{figure}

One can also calculate the Green function with retarded and advanced boundary conditions. In the retarded case, the loop integrals going into the vacuum polarization need to be calculated using the in-in formalism. This has been done in Ref.~\cite{Donoghue:2014yha}. The result is the same functional dependence, but with a different $i\epsilon $ prescription. In particular, the logarithm becomes
\beq
\log\left(-\left[(q_0+i\epsilon)^2-\vec{q}^2\right]\right)= \log\left(-q^2 -i\epsilon q_0\right)
= \log|q^2| -i\pi \theta(q^2)\left(\theta(q_0)-\theta(-q_0)\right)   \ \ .
\eeq
This shifts the location of the poles to the positions indicated in Fig. \ref{contour2}.

For $t>0$ we pick up only the usual
massless poles
\beq
D_{\textrm{ret}}(t>0,\vec{x}) = D_{\textrm{ret}}^{(0)}(t>0,\vec{x})
\eeq
However, here the unusual feature is that there is a contribution to the retarded Green function also for $t<0$,
\beq
D_{\textrm{ret}}(t<0,\vec{x})\equiv D_{\textrm{ret}}^<(t,\vec{x})
=  i\int \frac{d^3q}{(2\pi)^3 }\left[ \frac{e^{-i(E_q t -\vec{q}\cdot\vec{x})}}{2(E_q +i\frac{\gamma}{2E_q})} e^{-\frac{\gamma |t|}{2E_q}}- \frac{e^{i(E_q t -\vec{q}\cdot\vec{x})}}{2(E_q -i\frac{\gamma}{2E_q})} e^{-\frac{\gamma |t|}{2E_q}}\right]  \ \ .
\eeq
This also contains decaying exponentials. If we choose to use this as a classical Green function giving the response to an external source,
\beq
h_{\mu\nu}(t,x) = \int d^3x' \left[\int_{-\infty}^t dt' D_{\textrm{ret}}^{(0)}(t-t',x-x')
+ \int^{\infty}_t dt' D^<_{\textrm{ret}}(t-t',x-x')\right] J_{\mu\nu}(t', x')
\eeq
it would correspond to the propagation of the effect backwards in time. This is related to the microcausality violation on scales of order of the resonance width, known to be a feature of Lee-Wick type theories.

Finally, let us briefly comment on the formulation of our results in curved spacetime. One way to proceed is to consider the usual local momentum representation of the Feynman propagator, which employs Riemann normal coordinates~\cite{Birrell,Parker}. But the straightforward expansion of the action is not enough. Here the loop diagram is also an essential ingredient since it generates the width for the ghost state. The associated expansion in curvature of this result can be relatively complicated. Indeed, one of us has calculated the full one-loop diagram for the photon propagator~\cite{DB:2015a,DB:2015b}. Such considerations can be carried out for Lee-Wick type theories, but that would take us a bit far from the scope of the present work. We plan to return to this topic in a future publication.

\section{Cuts with stable and unstable particles}\label{explicitcuts}

Unitarity is realized by including only the stable particles in the unitarity sum. Because the all-orders proof of this including the unstable ghost, given in the next section using the methods of Veltman, is somewhat contrived and non-intuitive, we here give some simple examples to explore how this works. These examples also are useful in the discussion of the narrow-width approximation, which follows in Sec. \ref{NWA}

In calculating the discontinuity of a process or a Feynman diagram, one cuts the diagram(s) in all possible ways. For the propagators in the cut, one replaces the propagator by an on-shell delta function
\beq
\frac{i}{q^2-m^2+i\epsilon} \to 2\pi \delta(q^2-m^2)\theta(q_0)  \ \ .
\eeq
One also replaces all propagators on the far right-hand side of the cut (typically represented by a shaded region) by their complex conjugate\footnote{Not all textbooks include this latter rule, although it is important in multi-loop diagrams.}. This amounts to replacing $+i\epsilon$ by $-i \epsilon$,
\beq
\frac{i}{q^2-m^2+i\epsilon}  \to \frac{-i}{q^2-m^2-i\epsilon}
\eeq
for these propagators.

In Ref. \cite{Donoghue:2018izj}, we discussed the scattering in the spin-two partial wave in quadratic gravity and found it to be unitary. Specifically, we found
\beq\label{t2amplitude}
T_2(s) = - \frac{N_{\textrm{eff}} s}{640 \pi}\,\bar{D}_2(s).
\eeq
This is normalized such that in order to satisfy elastic unitarity we must have $\textrm{Im} T_2 = |T_2|^2$. This is satisfied when the amplitude has the form
\beq
T_2(s) = \frac{A(s)}{f(s) -i A(s)}= \frac{A(s)[f(s)+i A(s)]}{f^2(s)+A^2(s)}
\eeq
for any real functions $f(s),~A(s)$. This relation is satisfied with
\beq
A(s) =  - \frac{N_{\textrm{eff}} s}{640 \pi}.
\eeq
Because this example is essentially just scattering through the spin-two propagator, we start with the calculation of the discontinuity in propagators.

\subsection{Two-particle cut}

\begin{figure}[htb]
\begin{center}
\includegraphics[height=30mm,width=100mm]{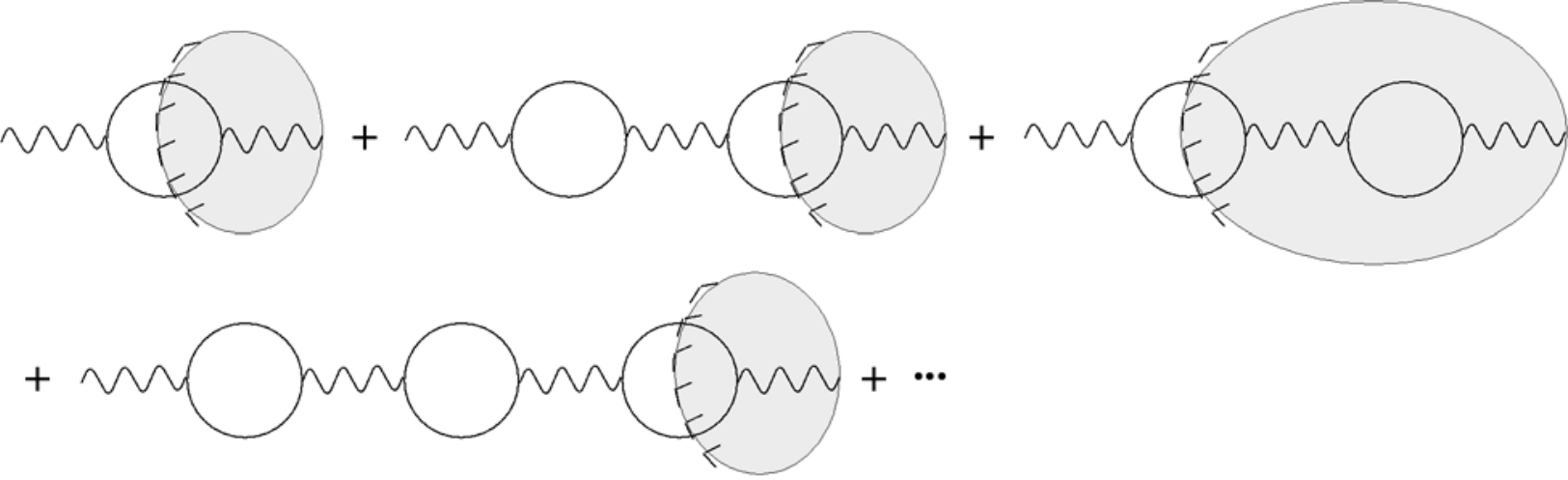}
\caption{The two particle cuts in the propagator. }
\label{twopart}
\end{center}
\end{figure}

For simplicity, we here study a scalar propagator, using the scalar proxy for quadratic gravity, Eq. \ref{scalargravity}. The vacuum polarization enters the propagator through the bubble sum. The cuts available in the bubble sum are shown in Fig. \ref{twopart}. The calculation of the basic cut is straightforward,
\beq
\textrm{Disc}_2 ~\Sigma (q) = \frac{\kappa^2 q^4(N+1)}{2}  \int \frac{d^4 k}{(2\pi)^4}~  2 \pi \delta(k^2)\theta(k_0) ~ 2\pi \delta ((q-k)^2) \theta((q-k)_0)  \ \ .
\eeq
The result is
\beq
\textrm{Disc}_2~ \Sigma (q^2)  = \frac{(N+1) \kappa^2q^4}{32\pi}
\eeq
Here the $(N+1)$ factor counts the number of massless states that the resonance can decay into.

We can also repackage the discontinuity in terms of the two-body decay width. This can be done by inserting a factor of
\beq
\int \frac{d^4 p}{(2\pi)^4}~ (2\pi)^4 \delta^4 (q-k-p) ~=~1
\eeq
and identifying the matrix element ${\cal M}_2 = \kappa q^2$, to write
\beq
\textrm{Disc}_2 ~\Sigma (q) = \frac{(N+1)}{2}  \int \frac{d^4 k}{(2\pi)^4}~ \frac{d^4 p}{(2\pi)^4}~ (2\pi)^4 \delta^4 (q-k-p) |{\cal M}_2|^2 2 \pi \delta(k^2)\theta(k_0) ~ 2\pi \delta ((q-k)^2) \theta((q-k)_0)  \ \ .
\eeq
The on-shell delta functions can be used to do the integration over the energy variables, leaving
\beq
\textrm{Disc}_2 ~\Sigma (q) = \frac{(N+1)}{2}  \int \frac{d^3 k}{(2\pi)^3 2\omega_k}~ \frac{d^3 p}{(2\pi)^3 2\omega_p}~ (2\pi)^4 \delta^4 (q-k-p) |{\cal M}_2|^2  \ \ .
\eeq
This latter relation can be recognized as the two-body decay width at the overall center of mass momentum $q$, up to an initial normalization factor of $1/2q$,  resulting in
\beq
\textrm{Disc}_2 ~\Sigma (q) = 2 q \Gamma_2(q)
\eeq
which is another statement of the unitarity relation.

When we take cuts in diagrams with multiple bubbles, the bubbles on each side of the cut can be summed to yield the dressed propagator. Recall that on shaded regions, the bubbles are calculated with a
$-i\epsilon$ prescription, which results in the complex conjugate of the usual bubble diagram. Performing this bubble sum, the result is
\beq\label{propagatorcut}
\textrm{Disc} ~D(q) = D(q)~ 2 q \Gamma_2(q) ~D^*(q) = - 2~ \textrm{Im}[D(q)]
\eeq
with
\beq\label{scalarproxy}
D(q) = \frac{1}{q^2 -\frac{\kappa^2 q^4}{2\xi^2(\mu)} -\frac{(N+1) \kappa^2 q^4}{32\pi^2}[\log (q^2/\mu^2)-i\pi]}
\eeq
which is the expected discontinuity relation for the propagator.

\subsection{Three particle cut}

\begin{figure}[htb]
\begin{center}
\includegraphics[height=15mm,width=90mm]{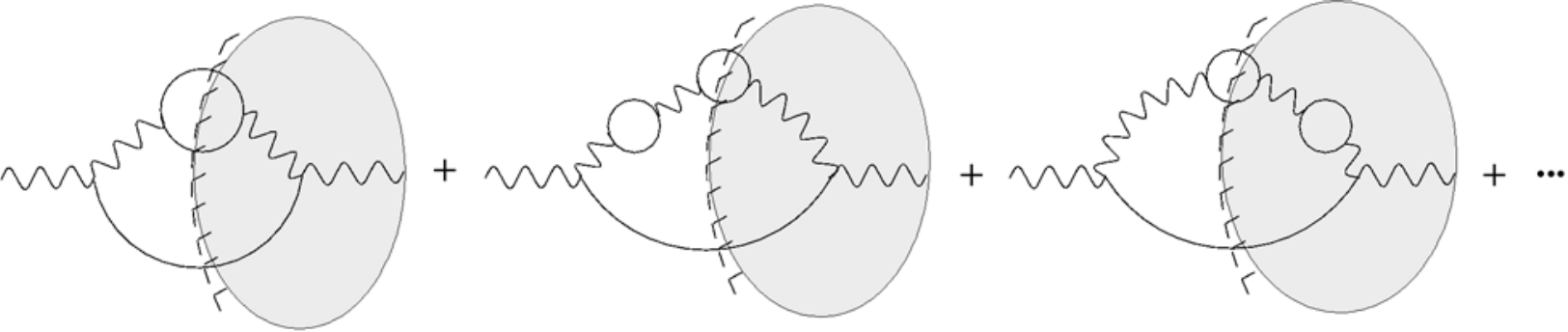}
\caption{The three particle cuts in the propagator. }
\label{threepart2}
\end{center}
\end{figure}

More interesting is the three particle cut, such as that shown in Fig. \ref{threepart2}. Here the basic cut is in a two loop integral.
\bea
\textrm{Disc}_3 ~ \Sigma(q) &=& \kappa^2 q^4 \int  \frac{d^4 k_1}{(2\pi)^4}~ \frac{d^4 k_2 }{(2\pi)^4}\frac{1}{(q- k_1)^2- \frac{\kappa^2 (q-k_1)^4}{2\xi^2}} \frac{(N+1)\kappa^2 (q-k_1)^2}{2}
\nn\\
&\times& ~2 \pi \delta(k_1^2)\theta(k_{10}) ~ 2\pi \delta (k_2^2)\theta(k_{20})~  2\pi \delta ((q-k_1-k_2)^2) \theta((q-k_1-k_2)_0) \frac{1}{(q- k_1)^2- \frac{\kappa^2 (q-k_1)^4}{2\xi^2}}     \ \ .
\eea

Here again there are bubble diagrams on each side of the cut, and these can be summed to yield the dressed propagator, with the propagators on the shaded regions being complex conjugated. We can package this more compactly, using the matrix element
\beq
{\cal M}_3 = \kappa q^2 \kappa (q-k_1)^2 D(q-k_1)
\eeq
with the propagator $D$ being given by Eq. \ref{scalarproxy}. Using this we have
\beq
\textrm{Disc}_3 ~ \Sigma(q) = \frac{N+1}{2} \int  \frac{d^4 k_1}{(2\pi)^4}~ \frac{d^4 k_2 }{(2\pi)^4}|{\cal{M}}_3|^2  ~2 \pi \delta(k_1^2)\theta(k_{10}) ~ 2\pi \delta (k_2^2)\theta(k_{20})~  2\pi \delta ((q-k_1-k_2)^2) \theta((q-k_1-k_2)_0)      \ \ .
\eeq
We can add an extra integration and an extra delta function by using momentum conservation to write $k_3=q-k_1-k_2$, such that
\beq
\textrm{Disc}_3 ~\Sigma(q) =\frac{N+1}{2}  \int  \frac{d^4 k_1}{(2\pi)^4}~ \frac{d^4 k_2 }{(2\pi)^4}\frac{d^4 k_3}{(2\pi)^4} ~(2\pi)^4 \delta^4(q-k_1-k_2-k_3)
|{\cal{M}}_3|^2  ~2 \pi \delta(k_1^2)\theta(k_{10}) ~ 2\pi \delta (k_2^2)\theta(k_{20})~  2\pi \delta ((k_3)^2) \theta((k_{30})      \ \ .
\eeq
Now the on-shell delta functions can be performed by doing the energy integral in each case,
\beq
\textrm{Disc}_3 ~ \Sigma(q) = \frac{N+1}{2}\int  \frac{d^3 k_1}{(2\pi)^3 2\omega_1}~ \frac{d^3 k_2 }{(2\pi)^3 2\omega_2 }\frac{d^3 k_3}{(2\pi)^3 2\omega_3} ~(2\pi)^4 \delta^4(q-k_1-k_2-k_3)
|{\cal{M}}_3|^2        \ \ .
\eeq
At this stage we recognize the formula for the three body decay  width $\Gamma_3$, such that
\beq\label{threepart}
\textrm{Disc}_3 ~\Sigma (q)= 2 q \Gamma_3(q)
\eeq
which is the desired unitarity relation.

\begin{figure}[htb]
\begin{center}
\includegraphics[height=20mm,width=30mm]{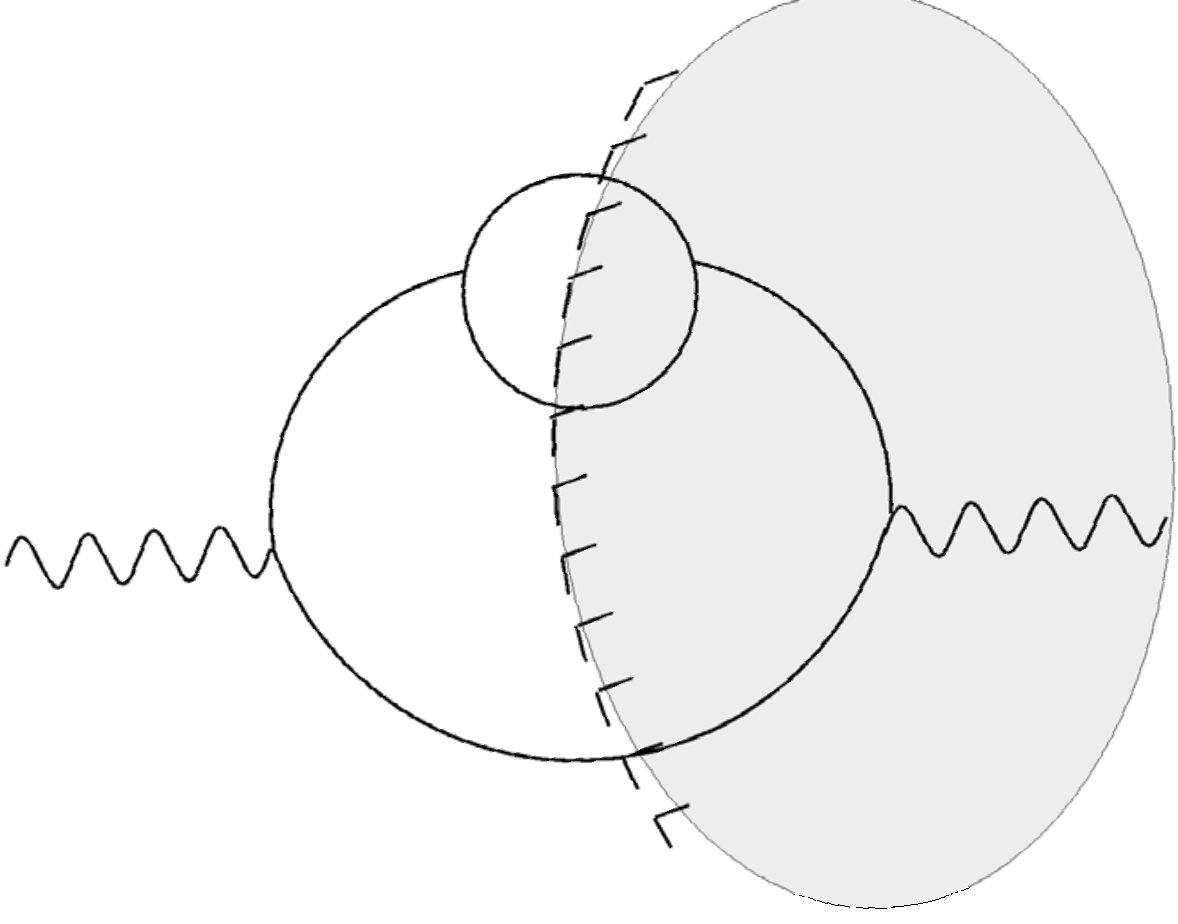}
\caption{Another three particle cut in the self energy function. }
\label{extracut}
\end{center}
\end{figure}

Besides the cut which we just calculated there is another one which does not involve the ghost in any way. It is shown in Fig. \ref{extracut}. It is simple to evaluate and just gives an extra term in the matrix element ${\cal M}$.

\subsection{Heuristic proof of unitarity}

The above calculations allow us to get a feel for why such theories satisfy unitarity even beyond these simple examples. We have seen that unitarity works with the stable particles as external states in the unitarity sum. The ghost resonance does not occur as an external state. This is the same pattern as normal resonances and the above calculation could be performed identically for normal resonances.

We also know from the analysis of subsection \ref{generic} that normal resonances and ghost resonances can be described in the same propagator using the coupling to the stable states described by the same $\Sigma(q)$.

Finally, we know from Veltman's work that normal resonances satisfy unitarity to all orders. It then follows that any discontinuity calculated with normal resonances in the intermediate states, can be converted into a discontinuity with ghost resonances by using  Eq. \ref{generalcase}. If the normal resonance satisfies the unitarity relation, the ghost resonance will also.

The above heuristic argument appears to be correct in direct calculations. Comparison with the formal proof of Veltman however superficially appears to indicate some differences between the two cases. This has to do with the differences with energy flow in the different parts of the propagators, a feature that we have explored in Sec. \ref{stability}. We will see that this difference can be accounted for and unitarity holds. We turn to this task next.

\section{Proof of unitarity with unstable ghosts }\label{Veltman}

In this section we give a proof that theories with unstable ghosts are unitary to all orders. The demonstration follows Veltman's argument~\cite{Veltman:1963}. For such purposes, we will derive the so-called cutting equation which implies unitarity, as we will discuss. The cutting equation contains terms associated with the imaginary part of propagators, given by the cut propagators. Essentially, the cutting equation is to be identified with the generalized optical theorem associated with an arbitrary process
$a \to b$:
\beq
i {\cal M}(a \to b) - i {\cal M}^{*}(b \to a)  =  - \sum_{f} \int d\Pi_{f} {\cal M}^{*}(b \to f) {\cal M}(a \to f)
(2\pi)^{4} \delta^{4}(a-f)
\label{opt_theo}
\eeq
which is just Eq.~(\ref{unit}) with the definition $\langle f |T| i \rangle = (2\pi)^{4}
\delta^{4}(p_{i} - p_{f}) {\cal M}(i \to f)$. The quantity $d\Pi_{f}$ is the Lorentz-invariant phase space~\cite{Schwartz:13} and the sum runs over all possible sets $f$ of intermediate states (single and multiparticle states) and there is an overall delta function assuring energy-momentum conservation. In the above expression, the ${\cal M}$'s are the invariant scattering matrix elements which are the non-trivial part of the $S$-matrix. On the other hand, the usual physical interpretation of the LSZ formula is that the S-matrix projects out one-particle asymptotic states from the time-ordered product of fields~\cite{Schwartz:13}. Since the spectral function contains contributions from multiparticle intermediate states with a continuous mass spectrum, this implies that the contribution coming from the cut propagators in the cutting equation should only contain the terms associated with the one-particle states. This also implies that the contributions coming from unstable particles are expected to be absent in Eq.~(\ref{opt_theo}). This is what we wish to prove here for models with unstable ghosts.

\subsection{Lehmann representations}

We begin by establishing our assumptions and some basic definitions. We assume that an expression for the propagator in terms of a Lehmann representation \cite{Kallen:1952zz, Lehmann:1954xi} is available, that is
\beq
G(p^2) = \frac{1}{\pi} \int_{0}^{\infty} ds \frac{\sigma(s)}{p^2 - s + i\epsilon}.
\label{lehmann}
\eeq
for stable particles, and
\beq
\widetilde{G}(p^2) = \frac{1}{\pi} \int_{0}^{\infty} ds \frac{\widetilde{\rho}(s)}{p^2 - s + i\epsilon}.
\label{lehmann2}
\eeq
for normal unstable particles. As remarked above we neglect any masses of stable particles.

For stable particles, there should be an isolated delta function contribution due to one-particle states:
\beq
\sigma(s) = {\cal Z} \pi \, \delta(s) + \rho(s).
\label{ss}
\eeq
On the other hand, for unstable particles the ``one-particle pole" is right on the branch cut; interactions essentially move the ``pole" off the branch cut, but at the cost of introducing an imaginary part into it. In this case one does not verify the presence of an asymptotic state according to the LSZ reduction formula and it is in this sense that such a particle is said to be unstable: There cannot be a isolated delta-function contribution to the spectral function since there are no one-particle states in the usual sense. In this case, in the narrow-width approximation the spectral function has a Breit-Wigner shape:
\beq
\widetilde{\rho}(s) \approx \frac{M \Gamma}{(s-M^2)^2 + M^2 \Gamma^2}
\label{su}
\eeq
where $\Gamma$ is the width of the resonance. For the normal unstable particle, the Lehmann representation has $\widetilde{\rho}(s) > 0$ and $\epsilon > 0$.

For theories with unstable ghosts, such as Lee-Wick QED and our scalar proxy for quadratic gravity given by Eq. \ref{scalargravity}, as discussed above the positive energy flow associated with the ghost propagates backwards in time. Moreover, the residue at the pole associated with the unstable ghost is always negative and the sign of the width is always opposite from normal resonances. For such reasons, for the ghost field we propose the following Lehmann representation
\beq
\widetilde{G}_{\textrm{GH}}(p^2) = - \frac{1}{\pi} \int_{0}^{\infty} ds
\frac{\widetilde{\rho}(s)}{p^2 - s - i\epsilon}.
\label{lehmannLW}
\eeq
This amounts to considering a ``anti time-ordered product" in the definition of the propagator. This full spectral representation for the LW propagator can be easily obtained from the modified Lehmann representation presented in Refs.~\cite{Grinstein:2008bg,DM:19} -- one just needs to employ a spectral function with a Breit-Wigner shape (in the narrow-width approximation) for both complex-conjugate poles.

In setting up the above Lehmann representations, we are clearly resorting to the partial fraction relation given by Eq.~(\ref{partialfrac}) for the case of the unstable ghost. We use this equivalent form for convenience and also for clarity of presentation. In any case, it is important to bear in mind that the stable particle and the ghost-like resonance are both parts of the same propagator, so the consideration of both entities as two separate particles cannot be pivotal to the procedure undertaken here. As such, all the arguments and results obtained in this section can be generalized for a single propagator such as the one given by Eq.~(\ref{generalcase}).

In the Lehmann representation for the stable particle presented above, ${\cal Z} = 1$ for the free theory. In such a case, field operators can only create a single particle from the vacuum. Hence, the contributions from multiparticle intermediate states, encoded in the spectral functions $\rho(s)$ and
$\widetilde{\rho}(s)$, are only present in the full interacting theory. In other words, the information about interactions are contained in ${\cal Z}$ as well as in the spectral functions. This implies that the Lehmann representation for unstable particles will never contain a ``free-field" contribution. We will return to this topic in due course.

\subsection{Cut propagators}

In what follows, an important ingredient in our discussion of unitarity is the decomposition of the propagator into terms with forward and backwards propagation in time. For the stable particle, one gets
\beq
i G(x-x') = \Theta(x_0-x_0') G^{+}(x-x') + \Theta(-x_0+x_0') G^{-}(x-x')
\eeq
where, using the spectral representation
\beq
G^{\pm}(x-x') =  \int \frac{d^{4} p}{(2\pi)^{3}} e^{-i p \cdot (x-x')} \theta(\pm p_0)
\frac{\sigma(p^2)}{\pi}
\eeq
The Green's functions $G^{\pm}$ are also known as cut propagators. Using that
\beq
i G^{\pm}(x-x';m^2) =  \pm \, 2\pi \int \frac{d^{4} p}{(2\pi)^{4}} e^{-i p \cdot (x-x')}
\theta(\pm p_0) \delta(p^2-m^2)
 \eeq
one can also write
\beq
G^{\pm}(x-x') = \pm \frac{1}{\pi} \int_{0}^{\infty} ds \, \sigma(s) \, i G^{\pm}(x-x';s).
\eeq
Hence, in momentum space
\beq
G^{\pm}(p^2) = 2\pi \theta(\pm p_0) \int_{0}^{\infty} ds \, \delta(p^2 - s) \frac{\sigma(s)}{\pi}.
\eeq
For the normal unstable particle, one finds a similar representation in terms of cut propagators, namely
\beq
i \widetilde{G}(x-x') = \Theta(x_0-x_0') \widetilde{G}^{+}(x-x') + \Theta(-x_0+x_0') \widetilde{G}^{-}(x-x')
\eeq
where
\beq
\widetilde{G}^{\pm}(x-x') = \pm \frac{1}{\pi} \int_{0}^{\infty} ds \, \widetilde{\rho}(s) \, i G^{\pm}(x-x';s).
\eeq
In momentum space:
\beq
\widetilde{G}^{\pm}(p^2) = 2\pi \theta(\pm p_0) \int_{0}^{\infty} ds \, \delta(p^2 - s)
\frac{\widetilde{\rho}(s)}{\pi}.
\eeq
The case of unstable ghost-like resonance is more subtle since here energy flow is backwards, among other features discussed above. Following the prescription proposed above, we write
\beq
i \widetilde{G}_{\textrm{GH}}(x-x') = \Theta(x_0-x_0') \widetilde{G}_{\textrm{GH}}^{-}(x-x')
+ \Theta(-x_0+x_0') \widetilde{G}_{\textrm{GH}}^{+}(x-x')
\eeq
where
\beq
\widetilde{G}_{\textrm{GH}}^{\pm}(x-x') = \mp \frac{1}{\pi} \int_{0}^{\infty} ds \,
\widetilde{\rho}(s) \, i G^{\pm}(x-x';s).
\eeq
In momentum space:
\beq
\widetilde{G}_{\textrm{GH}}^{\pm}(p^2) = - 2\pi \theta(\pm p_0) \int_{0}^{\infty} ds \, \delta(p^2 - s) \frac{\widetilde{\rho}(s)}{\pi}.
\label{LWm}
\eeq
For all cases above, in view of the reality of the spectral functions one has $G^{\pm} = (G^{\mp})^{*}$. Hence one finds that
\bea
-i G^{*}(x-x') &=& \Theta(x_0-x_0') G^{-}(x-x') + \Theta(-x_0+x_0') G^{+}(x-x')
\nn\\
-i \widetilde{G}^{*}(x-x') &=& \Theta(x_0-x_0') \widetilde{G}^{-}(x-x')
+ \Theta(-x_0+x_0') \widetilde{G}^{+}(x-x')
\nn\\
-i \widetilde{G}_{\textrm{GH}}^{*}(x-x') &=& \Theta(x_0-x_0') \widetilde{G}_{\textrm{GH}}^{+}(x-x')
+ \Theta(-x_0+x_0') \widetilde{G}_{\textrm{GH}}^{-}(x-x').
\eea
%

\subsection{Largest time equation}

Let us construct the so-called largest time equation. Consider an arbitrary connected diagram with $n$ vertices. All such vertices carry a spacetime coordinate $x_{i}$. We represent the diagram by the function $F(x_{1},x_{2},\ldots, x_{n})$. As usual each line connecting any two vertices $x_{i}$ and $x_{j}$ in $F$ is associated with a propagator. Suppose there is a coordinate $x_{m}$ associated with a given stable particle with the largest time component.  Then clearly $i G(x_{m}-x_{j}) = G^{+}(x_{m}-x_{j})$ and $i G(x_{k}-x_{m}) = G^{-}(x_{k}-x_{m})$ for all $x_k,x_j \in F$. Now consider the operation of circling the vertices of the diagram. Feynman rules can be introduced in order to take into account such an action:

\begin{itemize}

\item A line joining two uncircled (circled) vertices is associated with $G$ ($G^{*}$);

\item A line connecting a circled (uncircled) vertex $x_{k}$ to an uncircled (circled) vertex $x_{i}$ is to be associated with $G^{+}(x_{k} - x_{i})$ ($G^{-}(x_{k} - x_{i})$);

\item For any circled vertex one associates an overall minus sign.

\end{itemize}

\begin{figure}[htb]
\begin{center}
\includegraphics[height=50mm,width=80mm]{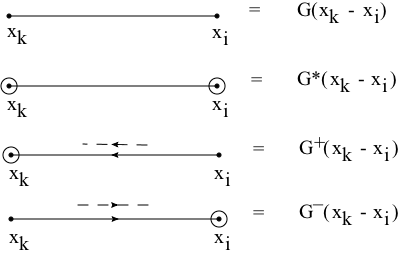}
\caption{Circling rules. The last two expressions are valid for stable particles and normal unstable particles. The arrow in the thick line connecting vertices represents the (positive) energy flow, whereas the arrow in the dashed line represents the flow of time.}
\label{circle}
\end{center}
\end{figure}

\begin{figure}[htb]
\begin{center}
\includegraphics[height=25mm,width=78mm]{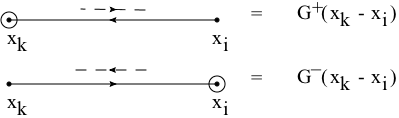}
\caption{Circling rules for the unstable ghost. Observe that energy flows in the opposite direction of the flow of time, in accordance with the description in the text.}
\label{circle2}
\end{center}
\end{figure}

For a thorough discussion, see Ref.~\cite{Veltman:74}. The first two rules are summarized in Fig.~\ref{circle}. To understand the third rule, consider, for instance, a generic diagram given by the expression
$$
F(x_{1},x_{2},x_{3}) = (ig)^{3} G(x_{1}-x_{2}) G(x_{2}-x_{3}) G(x_{3}-x_{1}).
$$
If we circle the vertices corresponding to $x_{2}$ and $x_{3}$, then the circling rules tell us that
$$
F(x_{1},\underline{x_{2}},\underline{x_{3}}) = (-ig)^{2}(ig) G^{-}(x_{1}-x_{2}) G^{*}(x_{2}-x_{3})
G^{+}(x_{3}-x_{1})
$$
where the coordinates related to the circled vertices were underlined. The $(-ig)^2$ in the above equation is the consequence of the usage of the third rule. Observe that one of the most important features of such rules is that energy flows from uncircled to circled vertices due to the presence of the theta function
$\theta(\pm p_{0})$ in the definition of $G^{\pm}(x-x';m^2)$, see Fig.~\ref{circle}. This feature is valid for all three cases, stable particle, normal unstable particles and unstable ghosts. The crucial difference is that for the latter the time flows in the opposite direction of the (positive) energy flow, see Fig.~\ref{circle2}

An immediate consequence of the above circling rules is that, if one considers an arbitrary circling of $F(x_{1},x_{2},\ldots, x_{n})$ and add to it the same resulting diagram but with the {\it opposite} circling of $x_{m}$, then the result is zero. In particular, if one considers all the $2^{n-1}$ possible circlings which keeps $x_{m}$, say, uncircled, and then sum up with all their possible counterparts which have $x_{m}$ circled, then the result is zero. But this is just the sum of all possible circlings of $F$:
\beq
\sum_{\textrm{circlings}} F(x_{1},x_{2},\ldots, x_{n}) = 0.
\label{LTE}
\eeq
This is the so-called largest time equation. Notice that in a theory containing unstable ghosts the Eq.~(\ref{LTE}) can only be found by using the largest time of the stable particles.

\subsection{The cutting equation}

\begin{figure}[htb]
\begin{center}
\includegraphics[height=25mm,width=80mm]{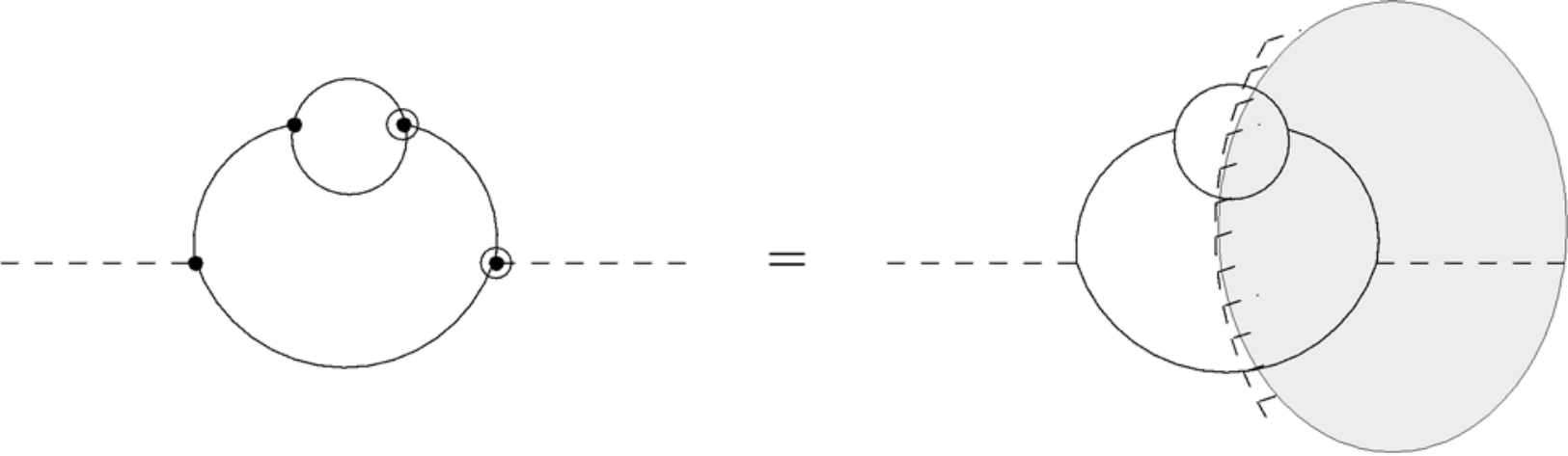}
\caption{A pictorial representation of a diagram with three cut lines. Dashed lines represent incoming and/or outgoing stable particles.}
\label{cut}
\end{center}
\end{figure}

Observe that the energy conservation in each vertex together with the energy flow dictated by the theta functions appearing in the cut propagators imply that many diagrams in Eq.~(\ref{LTE}) will be zero. In fact, the only non-vanishing diagrams are the ones in which circled vertices form connected regions that comprise one or more outgoing lines. For such non-vanishing diagrams one is allowed to drop the circles and mark such regions using a boundary line, which intersects one or more lines connecting uncircled and circled vertices. A typical example is illustrated in Fig.~\ref{cut}. The circling rules quoted above implies that such cut lines are represented by cut propagators $G^{+}$. Moreover, inside such connected regions (now represented by shaded parts in the diagram) all lines that were not cut should be represented by the complex conjugate of the propagator since they joined two circled vertices. In addition, all propagators associated with external lines connected to the diagram under consideration {\it inside} the shaded region should also be complex conjugated. Vertices inside such regions also carry an overall minus sign, on the account of the circling rules. Notice that the shaded region in Fig.~\ref{cut} satisfies a similar rule as compared to the shaded regions drawn in Figs.~\ref{twopart}--\ref{extracut} in the sense that the propagators inside them are complex conjugated.

In this procedure, only the fully uncircled diagram (denoted by $F(x_{1},x_{2},\ldots, x_{n})$) and the fully circled diagram (denoted by $\bar{F}(x_{1},x_{2},\ldots, x_{n})$) can both be drawn without a cut since they will not contain any cut propagators. Eq.~(\ref{LTE}) can thus be rewritten as
\beq
F(x_{1},x_{2},\ldots, x_{n}) + \bar{F}(x_{1},x_{2},\ldots, x_{n})
= - \sum_{\textrm{cuttings}} F(x_{1},x_{2},\ldots, x_{n}).
\label{cutting}
\eeq
This equation is also known as the cutting equation, or Cutkosky's cutting rule. Observe that energy is forced to flow towards the shaded region.

\subsection{Proof of unitarity to all orders}

Now we would like to compare Eqs.~(\ref{opt_theo}) and~(\ref{cutting}). First let us consider the left-hand sides of such equations. On the account of the circling rules, note that $\bar{F}$ is obtained from $F$ by complex conjugation of all propagators and multiplication by $-1$ of each vertex. In momentum space, this means that one must consider the replacement of every propagator with its complex conjugate in addition to reversing the direction of its momentum, $G(k) \to G^{*}(-k)$. In general, the usual Feynman rules attach an imaginary unit to each vertex (times possibly a coupling constant). From the third circling rule quoted above, one finds that every vertex in $\bar{F}$ should be complex conjugated if the coupling constants are real. The latter requirement is fulfilled if the Hamiltonian is Hermitian and contains only real fields. All such considerations imply that if $F$ is a diagram contributing to the process $a \to b$, then $\bar{F}$ describes the complex conjugate of the reversed process. In other words
\bea
\hspace{-4mm}
\sum_{\textrm{diagrams}} \Bigl[ F_{a \to b}(k_{1},k_{2},\ldots, k_{n})
+ \bar{F}_{a \to b}(k_{1},k_{2},\ldots, k_{n}) \Bigr]
&=& \sum_{\textrm{diagrams}} \Bigl[ F_{a \to b}(k_{1},k_{2},\ldots, k_{n})
+ F^{*}_{b \to a}(k_{1},k_{2},\ldots, k_{n}) \Bigr]
\nn\\
&=& i {\cal M}(a \to b) - i {\cal M}^{*}(b \to a)
\eea
where we summed over all possible diagrams contributing to the process $a \to b$, and $F(k_{1},k_{2},\ldots, k_{n})$ and $\bar{F}(k_{1},k_{2},\ldots, k_{n})$ are the Fourier transforms of $F(x_{1},x_{2},\ldots, x_{n})$ and $\bar{F}(x_{1},x_{2},\ldots, x_{n})$, respectively. On the other hand, if there are any complex fields, then the coupling constants can be complex. In this case one should also consider the complex conjugate of each interaction term, if one assumes that the Hamiltonian should be Hermitian. Again the circling rules will imply that $\bar{F}$ represents the complex conjugate of reversed processes~\cite{Rodenburg}.

Now let us compare the right-hand sides of Eqs.~(\ref{opt_theo}) and~(\ref{cutting}). Following Ref.~\cite{Rodenburg,Lang}, first consider the case in which only one internal line is being cut (we do not specify in particular whether such a line is associated with a stable or unstable particle). In this case, the diagram can be written as, in momentum space
\beq
F_{\textrm{single cut}}( a \to b ) = F_{\textrm{left}}(a \to k) G^{+}(k^2) \bar{F}_{\textrm{right}}(k \to b)
(2\pi)^{4} \delta^{4}(a-b)
\eeq
where $a$ collectively represents the sum of the momenta associated with the initial states and $b$ that of the final states. In addition, $G^{+}$ represents generically a cut propagator, not necessarily associated with a specific kind of particle (stable or unstable). In particular, for the unstable ghost one also has an overall minus sign in the expression for $G^{+}$, see Eq.~(\ref{LWm}). Notice that $\bar{F}_{\textrm{right}}$ corresponds to the part of the diagram inside the shaded region and hence it implies that all vertices in the region to the right of the cut are circled. The momentum $k$ of the cut line is equal to the total incoming momenta and outgoing momenta.

The diagrams $F_{\textrm{left}}(a \to k)$ and $\bar{F}_{\textrm{right}}(k \to b)$ cannot represent contributions to scattering matrix elements as they stand due to two main reasons:

\begin{enumerate}

\item The absence of the Lorentz-invariant phase space.

\item The momentum $k$ is not on-shell.

\end{enumerate}

By rewriting $G^{+}(k^2)$ as
\bea
G^{+}(k^2) &=& 2\pi \theta(k_0) \int_{0}^{\infty} ds \, \delta(k^2 - s) \frac{\sigma(s)}{\pi}
=   \int_{0}^{\infty} ds \, \frac{\sigma(s)}{\pi} \int \frac{d^4 q}{(2\pi)^{3}}
\, \theta(q_0) \delta(q^2 - s)  (2\pi)^{4} \delta^{4}(q-k)
\nn\\
&=& \int_{0}^{\infty} ds \, \frac{\sigma(s)}{\pi} \int \frac{d^3 q}{(2\pi)^{3}} \, \frac{1}{2\sqrt{{\bf q}^{2} + s}}
\, (2\pi)^{4} \delta^{4}(q-k) \Big|_{q^{0} = \sqrt{{\bf q}^{2} + s}}
\eea
one obtains that
\bea
F_{\textrm{single cut}}( a \to b ) &=& \int_{0}^{\infty} ds \, \frac{\sigma(s)}{\pi}
\int \frac{d^3 q}{(2\pi)^{3}} \, \frac{1}{2\sqrt{{\bf q}^{2} + s}}
\Bigl[ F^{*}_{\textrm{right}}(b \to q) F_{\textrm{left}}(a \to q)
\nn\\
&\times& (2\pi)^{4} \delta^{4}(a-q)  \Bigr] \Big|_{q^{0} = \sqrt{{\bf q}^{2} + s}} \, (2\pi)^{4} \delta^{4}(a-b).
\eea
The generalization to multiple cuts is straightforward~\cite{Rodenburg,Lang} (cf. Fig.~\ref{cut}):
\bea
F_{\textrm{cut}}( a \to b ) &=&
\left( \prod_{i} \int_{0}^{\infty} ds_{i} \, \frac{\sigma(s_{i})}{\pi}
\int \frac{d^3 q_{i}}{(2\pi)^{3}} \, \frac{1}{2\sqrt{{\bf q}_{i}^{2} + s_{i}}} \right)
\Bigl[ F^{*}_{\textrm{right}}(b \to \{q_{i}\}) F_{\textrm{left}}(a \to \{q_{i}\})
\nn\\
&\times& (2\pi)^{4} \delta^{4}\Bigl(a- \sum_{i} q_{i} \Bigr)  \Bigr] \Big|_{q_{i}^{0} = \sqrt{{\bf q}_{i}^{2} + s}} \, (2\pi)^{4} \delta^{4}(a-b).
\eea
We remark that here $\sigma(s)$ should be envisaged as representing the spectral function for any kind of particle, stable or unstable.

So we see that the Lorentz-invariant phase space emerges, but $q$ is {\it not} on-shell. This will only happen if the spectral function $\sigma(s)$ has a contribution from one-particle states. Otherwise $F_{\textrm{cut}}$ will not represent a contribution to the imaginary part of the scattering amplitude and hence will not contribute to the right-hand side of Eq.~(\ref{opt_theo}). From the Lehmann representations described above, we notice that this is the case only for stable particles. Unstable particles of any type, including the ghost-like resonances that appear in Lee-Wick models and quadratic gravity, {\it will not} contribute to unitarity sums. Unitarity is satisfied by the inclusion of only cuts from stable states in the unitarity sum. Finally observe that the sum over all possible cuttings is identical to the sum over all possible sets of intermediate states.

The fact that only stable states should enter in the unitarity sum can also be seen from a simple usage of the LSZ reduction formula~\cite{Schwartz:13}
\beq
\langle b |S| a \rangle = \left[ i \int d^{4} x_{1} e^{-i p_{1} \cdot x_{1}} (\Box_{1} + m^{2}) \right] \cdots
\left[ i \int d^{4} x_{n} e^{i p_{n} \cdot x_{n}} (\Box_{n} + m^{2}) \right]
\langle \Omega | T \{ \phi(x_{1}) \cdots \phi(x_{n}) \} | \Omega \rangle
\label{LSZ}
\eeq
associated with a quantum field $\phi(x)$ with mass $m$ and vacuum state $|\Omega\rangle$ of the full theory. By writing $S = 1 + i T$ and $\langle b | iT | a \rangle = (2\pi)^{4} \delta^{4}(\sum p) i {\cal M}$, one observes that the matrix element $i {\cal M}$ can be formally deduced from Eq.~(\ref{LSZ}). The exact $n$-point function will have the general structure~\cite{Schwartz:13}
\bea
\langle \Omega | T \{ \phi(x_{1}) \cdots \phi(x_{n}) \} | \Omega \rangle
&=& \int \frac{d^{4}q_{1}}{(2\pi)^{4}} e^{i q_{1} \cdot x_{1}} \cdots
\int \frac{d^{4}q_{n}}{(2\pi)^{4}} e^{-i q_{n} \cdot x_{n}}
i G(q_{1}^{2}) \cdots i G(q_{n}^{2})
\nn\\
&\times& {\cal D}(q_{1},q_{2},\ldots,q_{n})
(2\pi)^{4} \delta^{4}\Bigl( \sum_{i} q_{i} \Bigr)
\eea
where ${\cal D}(q_{1},q_{2},\ldots,q_{n})$ represents the sum of all amputated $n$-point diagrams. The exact propagators $G(q_{j}^{2})$ depicting the external legs can be written in terms of the Lehmann representations discussed above. On the other hand, the sum in Eq.~(\ref{opt_theo}) is over all possible sets of intermediate states, which may include states associated with unstable particles. However, when we use Eq.~(\ref{LSZ}) to calculate, say, the $S$-matrix elements $\langle f |S| a \rangle$ and then extract from it the associated scattering amplitude ${\cal M}(a \to f)$, one will get a non-vanishing contribution only if the propagators $G(q_{f}^{2})$ have a term associated with one-particle states. In particular, this implies that the right-hand side of Eq.~(\ref{opt_theo}) will not contain any contribution from intermediate unstable states.

There is one point that we have not discussed so far and that deserves a careful attention. All the above derivations were carried out for scalar fields. Does such a procedure generalize to more general fields carrying additional degrees of freedom? We know that, for the generalized optical theorem to hold in general, the numerator of a propagator must be equal to the sum over physical spin states~\cite{Schwartz:13}. So this sum must also be present in the numerator of the cut propagators in order to identify the right-hand sides of Eqs.~(\ref{opt_theo}) and~(\ref{cutting}). This is indeed the case for fermions, and one can easily decompose the fermionic propagator into positive- and negative-frequency parts~\cite{Veltman:74}. A Lehmann representation is also available~\cite{Schweber,Itzykson}. On the other hand, for photons the argument is more subtle. Indeed, the numerator of the propagator is not just the sum over physical polarizations~\cite{Schwartz:13}: Both have a distinct longitudinal part. But also both carry a
$\eta_{\mu\nu}$ factor. So, using the Ward identity for the scattering amplitudes (which contain a sum over physical polarizations) and gauge invariance for the propagator (which allows us to eliminate the longitudinal term by choosing, say, the Feynman gauge), then it is easy to show that unitarity holds. In particular, in the Feynman gauge the photon propagator can be decomposed  into positive- and negative-frequency parts~\cite{Rodenburg}. A Lehmann representation can also be derived for the photon propagator~\cite{Schweber,Itzykson}. In turn as above a full spectral representation for the associated ghots-like propagator can be easily obtained from the modified Lehmann representation in the Feynman gauge~\cite{DM:19}. The Ward identity in Lee-Wick gauge theory has also been demonstrated~\cite{Grinstein:2008gt}.

As one can note from the above discussion, the case of quadratic gravity deserves a careful treatment. Details concerning the Ward identity and the sum over physical spin states should be taken into account. Moreover, when considering an explicit expression for the spectral function one generally considers a resummation for internal propagators and then extracts the imaginary part of the resulting expression. But it is known that such a resummation procedure may violate the Ward identity if not performed with due care~\cite{Rodenburg}. Hence one must be careful in generalizing the procedure adopted here for the case of quadratic gravity. We hope to fully explore this issue in a future work.

\section{The narrow-width approximation and the Lee-Wick contour}\label{NWA}

Although the unitarity prescription is to include only cuts on the stable particles and not on the resonances, in practice we often ignore this when the resonances live long enough that we can treat them as stable in practical calculations. For normal resonances, this narrow widdth approximation (NWA) yields the correct answer in the limit that the width can be neglected. In this section we treat ghost-like resonances in this limit. We will see that we get a correct result if we take the $\Gamma \to 0 $ limit in the proper calculation, with stable particle cuts. However, if we were to treat the ghost as a stable particle from the start, we would get a different and incorrect answer. This can be rescued by a slight change of the integration contour, that proposed by Lee and Wick.

Here is how the NWA works. When studying loops with the resonance inside the loop, we can calculate the cuts using the usual cutting rules on the stable particles. We did this explicitly for two and three particle cuts in Sec. \ref{explicitcuts}. As seen in Eq. \ref{propagatorcut}, this yields
\beq
\textrm{Disc}~D(q) = D(q)~ 2q \Gamma(q)~ D^*(q) = \frac{2 q \Gamma(q)}{(q^2-m_r^2)^2 +(m_r \Gamma(q))^2}
\eeq
by the unitarity relation for the self-energy. In the limit that the width goes to zero, the contribution becomes vanishingly small except very close to the resonance. Using the representation of the delta function
\beq
\delta (x) = \lim_{\epsilon\to 0 } \frac{1}{\pi} \frac{\epsilon }{x^2+\epsilon^2}
\eeq
one finds that
\beq
\lim_{\Gamma \to 0} \textrm{Disc}~D(q) = 2\pi \delta(q^2-m_r^2)
\eeq
which reproduces the cutting rule for a stable particle. We note that, following from the work that we have presented above, this result is true for both normal resonances and ghost resonances. This implies that the usual cutting rules also apply to ghosts if we wish to approximate them as stable particles.

However, if one starts with the ghost field in a loop and calculates naively, one does not get the correct result. This can be demonstrated by exploring the three particle cut, which we calculated explicitly above. In this case, the NWA for the process consists of one massless field with the usual $i\epsilon$ structure plus one massive ghost field with a small width. Because the ghost is unstable and decays to two massless fields, the actual intermediate state will be three massless fields. We will show that this cut naively vanishes when calculated in the standard manner\footnote{There are many ways to obtain this result. Our presentation closely follows that of Schwartz~\cite{Schwartz:13}, and the reader can see there how the cutting rules arise for non-ghost fields.} .

\begin{figure}[htb]
\begin{center}
\includegraphics[height=70mm,width=90mm]{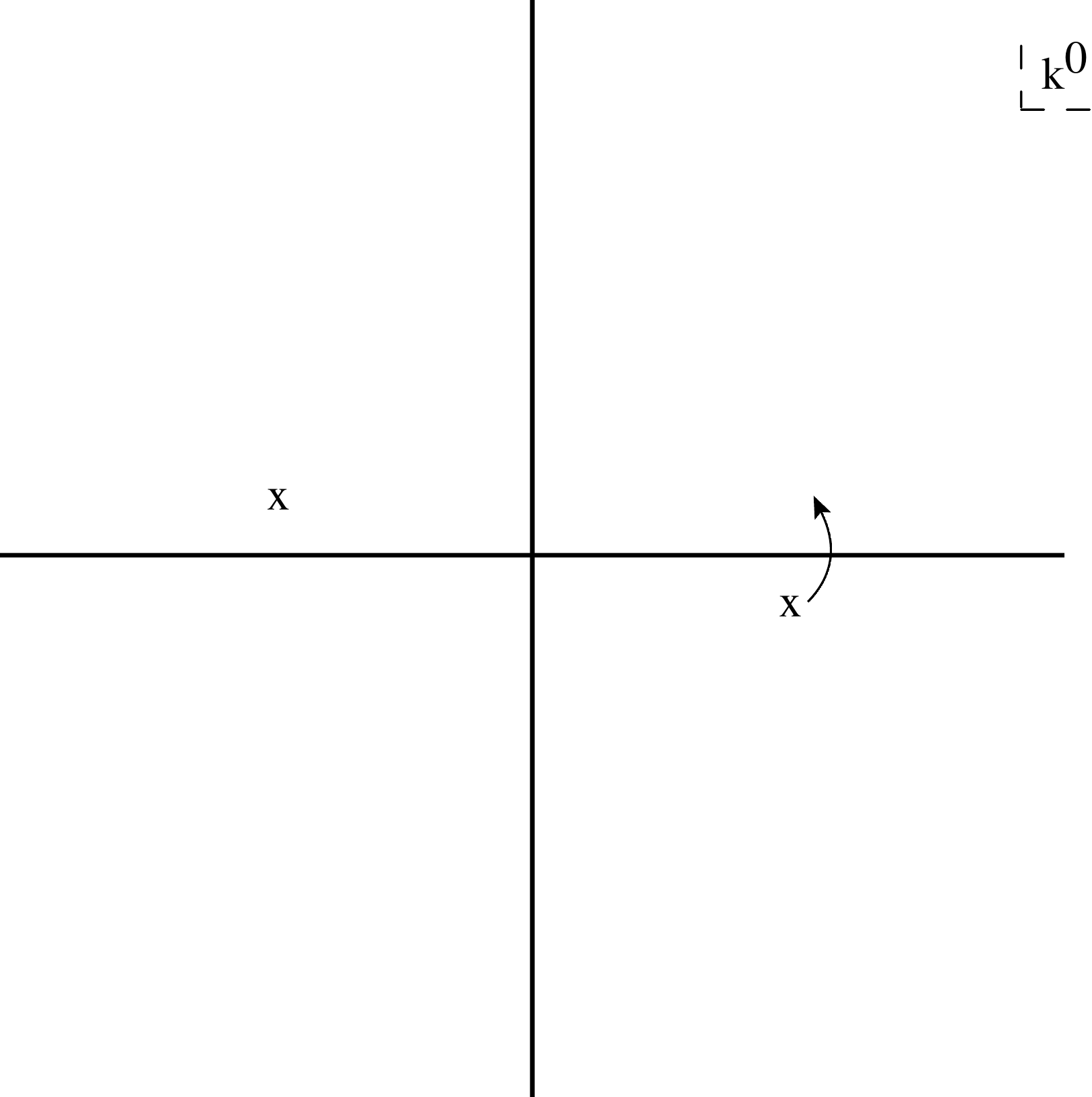}
\caption{The shift in pole location relating the Feynman propagator and the advanced propagator. }
\label{normal}
\end{center}
\end{figure}

\begin{figure}[htb]
\begin{center}
\includegraphics[height=70mm,width=90mm]{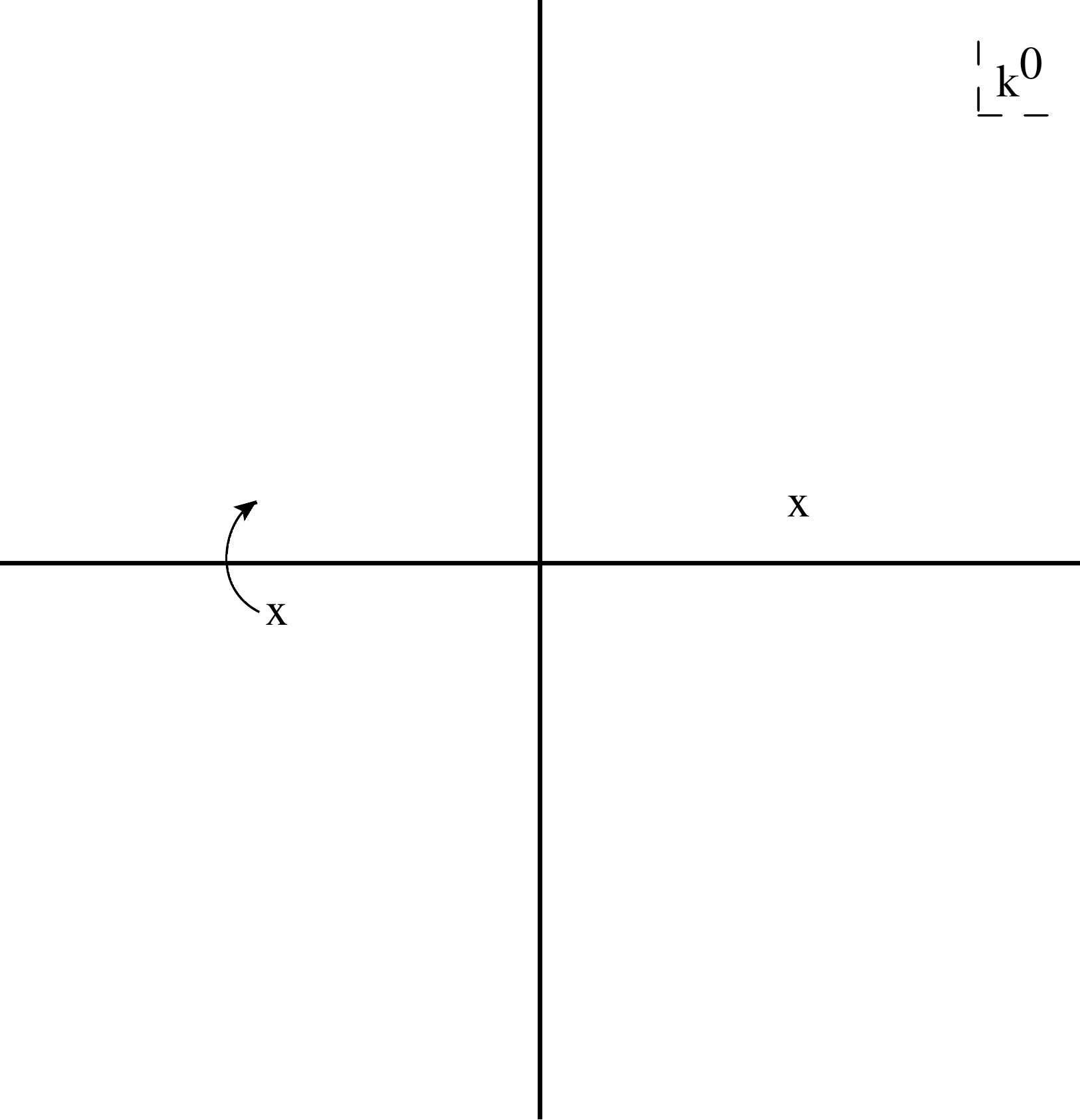}
\caption{The shift of a pole for the ghost particle. }
\label{ghostshift}
\end{center}
\end{figure}

Consider the amplitude with one massless field and one ghost
\beq
i {\cal M} =-  \int \frac{d^4 k}{(2\pi)^4} \frac{i}{(k-q)^2+i\epsilon}\frac{-i}{k^2 -m^2 -i\gamma}
\eeq
We will treat the width $\gamma$ as a small parameter. By shifting the poles in the normal Feynman propagator as illustrated in Fig. \ref{normal}, one write it in terms of the advanced propagator with poles only above the real axis via
\beq
\frac{i}{k^2+i\epsilon} = iD_A(k) +\frac{\pi}{\omega_k} \delta(k_0 -\omega_k)
\eeq
with $\omega_k= |{\bf{k}}|$, and
\beq
D_A =\frac{1}{2\omega_k}\left(\frac{1}{k_0-\omega_k-i\epsilon}- \frac{1}{k_0+\omega_k-i\epsilon} \right)
\eeq
Using the pole structure shown in Section \ref{stability},  one can do the equivalent transformation for the ghost propagator
\beq\label{ghosttrans}
\frac{-i}{k^2-m^2-i\gamma} = - iD_{``A''}(k) +\frac{\pi}{E_k} \delta(k_0  +E_{k-p})
\eeq
with $E_{k} = \sqrt{\bf{k}^2 +m^2}$, valid for infinitessimal $\gamma$, with
\beq
D_{``A''}(k) =\frac{1}{2E_k}\left(\frac{1}{k_0-E_k-i\gamma}- \frac{1}{k_0+E_k-i\gamma} \right) \ \
\eeq
using the shift of poles illustrated in Fig. \ref{ghostshift}. Here we have put the $A$ in quotes as one notes from the work in section \ref{stability} that this is not the actual ghost propagator with advanced boundary conditions. However, it has poles only above the real axis, and that is what is important in the present calculation.

\begin{figure}[htb]
\begin{center}
\includegraphics[height=70mm,width=90mm]{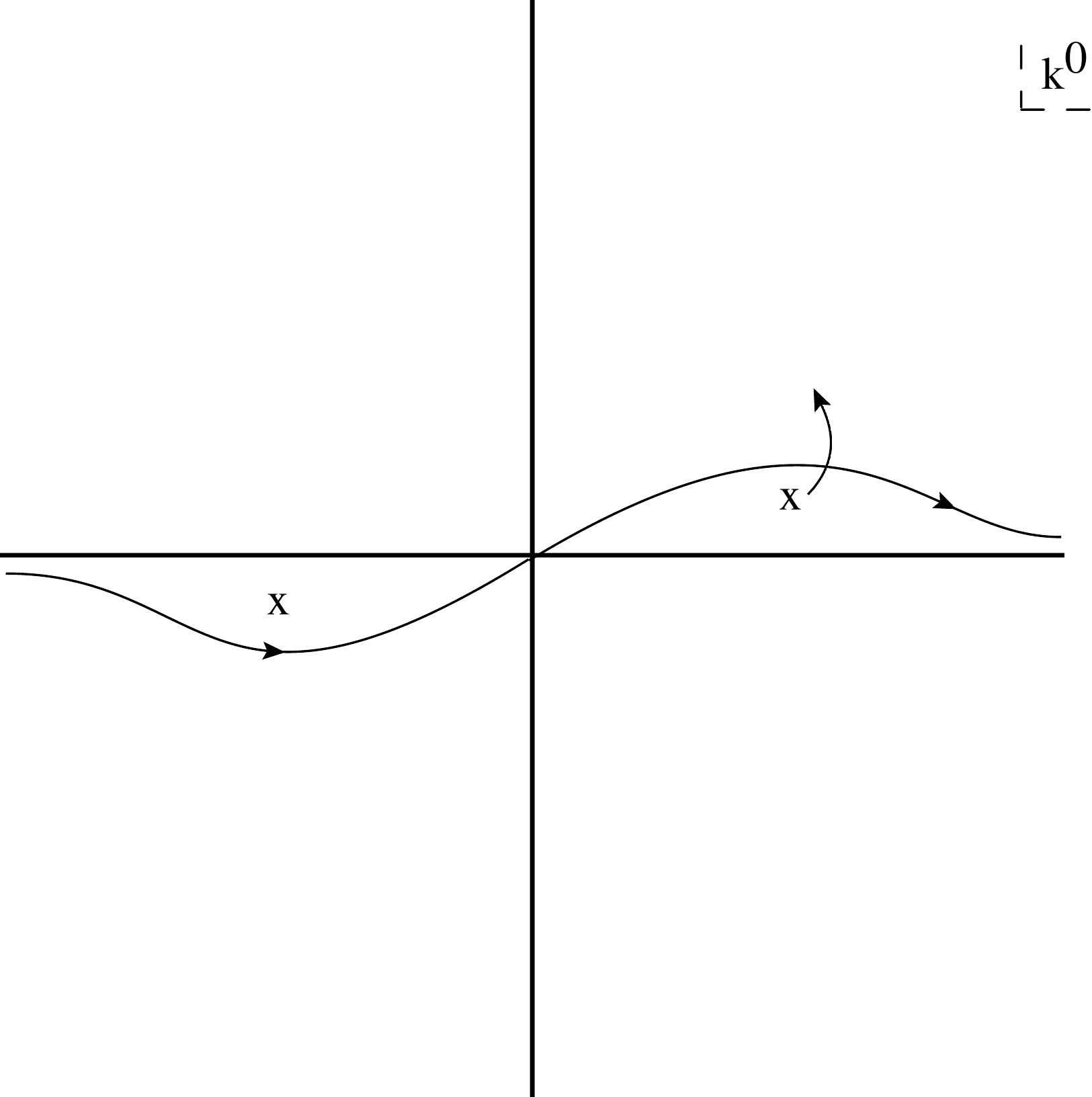}
\caption{Using the Lee-Wick contour. }
\label{withLW}
\end{center}
\end{figure}

The amplitude then turns into
\beq
i {\cal M} = -\int \frac{d^4 k}{(2\pi)^4} \left[iD_A(k-q) +\frac{\pi}{\omega_{k-q}} \delta(k_0- q_0 -\omega_{k-q})\right]\left[-i D_{``A''}(k) +\frac{\pi}{E_k} \delta(k_0 +E_k)
\right]
\eeq
If we work in the center of mass where $q$ is purely timelike, then $\omega_{k-q}=\omega_k=|{\bf k}|$. The contour can be closed below the real axis for the product of the two advanced propagators, with a vanishing result. The two delta functions cannot be simultaneously satisfied.  Only the cross-terms remain.

Having discarded the vanishing terms, the advanced propagators can be converted back into Feynman propagators, such that
\beq
{\cal M} = \int \frac{d^4 k}{(2\pi)^4} \left[D_F(k-q)~ \frac{\pi}{E_k} \delta(k_0 +E_k)+  D^g_{F}(k) \frac{\pi}{\omega_{k}} \delta(k_0- q_0 -\omega_{k})\right]
\eeq
Finally, we can pick out the imaginary part of this amplitude
\beq\label{impart}
{\rm Im}[{\cal M}] =- \int \frac{d^4 k}{(2\pi)^4} \left[   \pi \delta((k-q)^2) ~\frac{\pi}{E_k} \delta(k_0 +E_{k}) -\frac{\pi}{\omega_k} \delta(k_0 -q_0-\omega_k)~\pi \delta(k^2 -m^2)
\right]
\eeq
In both cases, the delta functions cannot be simultaneously satisfied. In the first term, we have $k_0 = - E_k$, while the second one would then require $(-\sqrt{{\vec k}^2 + m^2} -q_0)^2 = {\vec k}^2$, which is impossible. The second term also vanishes because $q_0> m$. Calculated this way, the imaginary part of the amplitude vanishes, in contrast with the direct calculation in Eq. \ref{threepart}.

We note that the problem comes from the $\delta (k_0+E_k)$ in the first term of this expression, which can be traced back to the same factor in Eq. \ref{ghosttrans}. For a normal particle this factor is $\delta (k_0-E_k)$, in which case both delta functions can be satisfied, yielding the usual rule for the cut. This problem is solved by a modification of the contour in doing the loop momentum integration, as originally described by Lee and Wick. The contour is chosen to run above the positive pole and below the negative one, as illustrated in Fig. \ref{withLW}. Using this contour, the shift in the pole location will now yield a factor of $\delta (k_0-E_k)$, and we get the usual cutting rule, as required.

This modification of the contour is introduced in order to reproduce the result of the correct discontinuity calculation presented in section \ref{explicitcuts}. It is needed because one has attempted to (improperly) take the cut across the unstable ghost as if it were a stable particle.

\section{Summary}

The ghost-like modes which occur in theories with quartic propagators require a re-evaluation of many of the basic properties of quantum field theory. A crucial observation is that these modes correspond to an unstable resonance, because of their coupling to lighter states in the theory. There are important consequences of this fact.  This removes these modes from the asymptotic spectrum, which consists only of stable particle states. In a path integral treatment, one does not have to perform free-field quantization for such states as they appear only in intermediate state propagators. We have shown that the energy flow is different than normal particles: What is usually claimed as positive energy propagates backwards in time. Nevertheless, the various Green functions imply stability in Minkowski space. The propagation in time is backwards from normal particles, and this contributes to a violation of microcausality over timescales of order the inverse width of these modes\footnote{We will discuss causality further in a separate paper \cite{DMcausality}.}.

Most importantly, building on the original work by Lee and Wick, we have given a proof that unitarity is satisfied to all orders in such theories, and provided explicit examples of how this occurs. Discontinuities are calculated by only including cuts on the stable particles of the theory. The proof follows the technique of Veltman, which was originally applied to reach the same conclusion about normal resonances. However the narrow-width approximation, in which cuts are applied to the unstable particles as if they were stable, appears differently for these resonances, and this can be accounted for in this limit by using the Lee-Wick contour. It should be understood that the Lee-Wick contour must be used only when one allows for the ghost as a particle in a loop (and this can only be done in the narrow-width approximation). Otherwise, the standard contour for the Feynman propagator is to be used.

An important point that should be highly appreciated is that the stable particle and the resonance appear in the same propagator, which means that eventually we are discussing two features of a single (quartic) propagator. The separation into two propagators (or the introduction of an auxiliary field at the level of the action) is to some extent artificial and it should be envisaged only as a convenient procedure in order to unveil in a manifest way the ghost feature of the propagator.

Overall, we see that these higher derivative theories are healthier than originally expected, as long as the microcausality violation occurs over short enough time scales that it is unconstrained by present experiments. While one can make higher derivative variants of any theory, the most important physical application is to gravity. The gravitational interaction requires higher derivatives in the action for the renormalizaton of the theory, and indeed quadratic gravity proves to be a renormalizeable theory of quantum gravity. It follows from our work here that it is also unitary and stable near flat space. It is in this regard the most conservative ultraviolet completion of quantum gravity, and deserves further exploration.

\section*{Acknowledgements} We would like to thank the following for useful comments or discussions: C. Burgess, A. Denner, B. Holdom, B. Holstein, P. Mannheim, D. O'Connell, M. Peskin, J. Ren, A. Salvio,  M. Schwartz, G. 't Hooft, A. Tolley, and C. Wetterich. The work of JFD has been partially supported by the US National Science Foundation under grant NSF-PHY18-20675. The work of GM has been partially supported by  Conselho Nacional de Desenvolvimento Cient\'ifico e Tecnol\'ogico - CNPq under grant 307578/2015-1 (GM) and Funda\c{c}\~ao Carlos Chagas Filho de Amparo \`a Pesquisa do Estado do Rio de Janeiro - FAPERJ under grant E-26/202.725/2018.

\end{document}